\documentclass[12pt,twocolumn,dutch,english,british]{iopart}
\usepackage[T1]{fontenc}
\usepackage[latin9]{inputenc}
\usepackage{babel}
\usepackage{textcomp}
\usepackage{graphicx}
\usepackage[unicode=true,
 bookmarks=true,bookmarksnumbered=false,bookmarksopen=false,
 breaklinks=false,pdfborder={0 0 1},backref=false,colorlinks=false]
 {hyperref}
\hypersetup{pdftitle={Reconnection and merging of positive streamers in air},
 pdfauthor={S. Nijdam}}

\makeatletter

\providecommand{\tabularnewline}{\\}

\usepackage{iopams}
\usepackage{setstack}

\usepackage{iopams}\newcommand{\SortNoop}[1]{}

\@ifundefined{showcaptionsetup}{}{%
 \PassOptionsToPackage{caption=false}{subfig}}
\usepackage{subfig}
\makeatother

\begin{document}

\title{Probing background ionization: Positive streamers with varying pulse
repetition rate and with a radioactive admixture}

\author{S~Nijdam$^{1}$, G Wormeester$^{2}$, E~M~van~Veldhuizen$^{1}$
and U~Ebert$^{1,2}$}

\address{$^{1}$ Eindhoven University of Technology, Dept.\ Applied Physics\\
 P.O. Box 513, 5600 MB Eindhoven, The Netherlands}

\address{$^{2}$ \foreignlanguage{dutch}{Centrum Wiskunde \& Informatica}\foreignlanguage{english}{
(CWI), Amsterdam, The Netherlands}}

\ead{s.nijdam@tue.nl}
\begin{abstract}
Positive streamers need a source of free electrons ahead of them to
propagate. A streamer can supply these electrons by itself through
photo-ionization, or the electrons can be present due to external
background ionization. Here we investigate the effects of background
ionization on streamer propagation and morphology by changing the
gas composition and the repetition rate of the voltage pulses, and
by adding a small amount of radioactive $^{85}$Kr.

We find that the general morphology of a positive streamer discharge
in high purity nitrogen depends on background ionization: at lower
background ionization levels the streamers branch more and have a
more feather-like appearance. This is observed both when varying the
repetition rate and when adding $^{85}$Kr, though side branches are
longer with the radioactive admixture. But velocities and minimal
diameters of streamers are virtually independent of the background
ionization level. In air, the inception cloud breaks up into streamers
at a smaller radius when the repetition rate and therefore the background
ionization level is higher. When measuring the effects of the pulse
repetition rate and of the radioactive admixture on the discharge
morphology, we found that our estimates of background ionization levels
are consistent with these observations; this gives confidence in the
estimates.

Streamer channels generally do not follow the paths of previous discharge
channels for repetition rates of up to 10\,Hz. We estimate the effect
of recombination and diffusion of ions and free electrons from the
previous discharge and conclude that the old trail has largely disappeared
at the moment of the next voltage pulse; therefore the next streamers
indeed cannot follow the old trail.
\end{abstract}

\submitto{\JPD}

\maketitle

\section{Introduction\label{sec:BG-ionization-Introduction}}

Positive streamers require a source of electrons in front of the streamer
head in order to propagate. These electrons can either be created
by non-local photo-ionization from the streamer head itself, or they
can already be present in the form of background ionization, either
freely or attached to an electronegative molecule like oxygen. Photo-ionization
is a strong non-local mechanism in air that depends on the presence
of both nitrogen and oxygen: fast electrons in the discharge excite
nitrogen levels that subsequently emit UV-photons that then can ionize
oxygen. However, in past work~\cite{Nijdam2010}, we have shown that
even in nitrogen with impurity levels well below 1 ppm, positive streamers
propagate at similar velocities as in air, but they are thinner, less
straight and they branch more; that even can resemble some feather-like
structure. The similarity of streamer velocities in air or high purity
nitrogen indicates that either low levels of oxygen still enable sufficient
photo-ionization for a streamer to propagate or that the streamers
can propagate by background ionization. Streamer simulations~\cite{Wormeester2010}
actually have shown that even an oxygen level of 1 ppm in nitrogen
can support streamer propagation by photo-ionization at velocities
similar to those in air, but that background ionization can support
streamer propagation as well, and that this mechanism becomes more
important at lower oxygen densities.

Here we focus on the effects of background ionization by varying its
level. Under normal circumstances, there are two probable sources
for background ionization: natural background ionization from cosmic
radiation or radioactivity, and remaining ionization from previous
discharges in a repetitive discharge mode. In the present experiments,
we again largely vary the nitrogen:oxygen ratio to modify photo-ionization,
and we now vary the repetition rate of the voltage pulse and we use
a radioactive admixture to change the background ionization.

In air, electrons produced by any mechanism quickly attach to oxygen
molecules. Therefore the ionization mostly consists of positive and
negative ions. At the moderate electric fields outside the streamer
head, these positive and negative ions are unable to create secondary
ionization (and therefore an avalanche). Only when the local electric
field is sufficiently high, e.g., close to a streamer head, an electron
can detach from a negative oxygen atom and it then can create an avalanche.
In nitrogen of sufficiently high purity the recombination rate of
electrons and ions can be higher than the attachment rate of electrons
to oxygen molecules and therefore most electrons will remain free
until recombination.\\

Natural background ionization levels by radioactivity and cosmic rays
are normally around 10$^{3}$--10$^{4}$\,cm$^{-3}$ in natural air
at standard temperature and pressure (Pancheshnyi~\cite{Pancheshnyi2005}
and references therein). The actual level depends on many parameters,
including building material type, ventilation, room or vessel dimensions
and pressure. In our stainless steel vacuum vessel, the ionization
levels will be lower than in ambient air at the same pressure ---
except, of course, with the radioactive admixture. This is because
we use artificial air and high purity nitrogen rather than ambient
air (although the supplier did not specify the radon levels), the
steel emits less radon than building materials and the vessel may
shield part of the cosmic rays. In practice, it is very difficult
to estimate the background ionization level inside the vessel, but
it can be safely assumed to be lower than 10$^{3}$\,cm$^{-3}$.

The level of background ionization caused by previous discharges in
a mode of repetitive voltage pulses depends on the pulse repetition
rate and on the streamer strength and density. At high streamer densities
(when the streamer channels fill most of the discharge volume) a 1\,Hz
repetition rate can create a background ionization level of about
10$^{7}$\,cm$^{-3}$ according to simple estimates by Pancheshnyi~\cite{Pancheshnyi2005}
as will be discussed in more detail in Section 2.

\subsection{Experimental results}

Hartmann and Gallimberti~\cite{Hartmann1975} have experimented with
positive streamer discharges in a jet so that rest products of previous
discharges are removed before the next discharge. They see a clear
effect of the jet on streamer properties, but attribute this to the
removal of nitrogen metastables and not of residual ionization.

Takahashi \emph{et al.}~\cite{Takahashi2011} have used a KrF laser
to produce UV-radiation of 248\,nm wavelength which produced enhanced
levels of background ionization in an atmospheric pressure argon discharge.
They observe that in a 30\,mm point-plane streamer discharge with
a 20\,kV pulse, branching is suppressed significantly when the background
ionization is above 5$\cdot$10$^{7}$\,cm$^{-3}$. A similar result
was found by Mathew \emph{et al.}~\cite{Mathew2007} who have studied
the effects of preionization (by x-ray radiation) on the uniformity
of discharges in F$_{2}$ excimer laser gas mixtures (He/F$_{2}$
mixtures with various ratios). They found that at preionization levels
of 10$^{7}$\,cm$^{-3}$\,bar$^{-1}$ and higher the discharge was
diffuse (uniform) and at lower preionization levels, the discharge
became filamentary or streamer-like.

\subsection{Theoretical results}

The older literature~\cite{Dyakonov1989,Raizer1998a} offers a simple
argument for why the exact source of electrons ahead of the streamer
should not influence its properties too much, if electrons are available
at all: as the electrons are exponentially multiplied in the active
ionization region ahead of the streamer, where the electric field
is above the breakdown value, the precise electron density at the
front edge of the active zone should influence streamer properties
only logarithmically.

Probably inspired by this argument, in early streamer simulations~\cite{Dhali1987,Vitello1994}
the photo-ionization in air was replaced by background ionization
to ease the computational demands. Dhali and Williams~\cite{Dhali1987}
used a high uniform background ionization level of the order of 10$^{5}$--10$^{8}$\,cm$^{-3}$,
while Vitello \textit{et al.}~\cite{Vitello1994} used a more realistic
value of 10$^{3}$\,cm$^{-3}$. Later, appropriate numerical means
were made available to simulate photo-ionization, as reviewed, e.g.,
in~\cite{Luque2011}.

The effects of different background ionization levels on streamers
were simulated recently by Bourdon \emph{et al.~}\cite{Bourdon2010}
and by Wormeester \emph{et al.}~\cite{Wormeester2010}. Bourdon \emph{et
al.} simulated both positive and negative streamers in preheated air
where pulse repetition rates of 10--30\,kHz were approximated by
homogeneous background ionization densities of up to 10$^{9}$\,cm$^{-3}$
(see Section~2 for such estimates). They found that the background
ionization density has a small influence on negative streamers, but
a significant influence on positive streamers; this agrees with earlier
results by Luque \textit{et al.}~\cite{Luque2008} where the same
asymmetry was found for the effect of photo-ionization on positive
or negative streamers, and an explanation was given.

Wormeester \emph{et al.} found similar results and studied also different
nitrogen:oxygen ratios. In that work, we show that photo-ionization
and background ionization have roughly the same effect on streamer
velocities, and that only a low contribution from either of these
processes is required to ensure streamer propagation (e.g., background
ionization larger than 10$^{3}$\,cm$^{-3}$). For example, a concentration
of 1\,ppm (parts per million) oxygen in nitrogen is enough to make
a streamer propagate through photo-ionization, without the need for
background ionization. In this case, the propagation velocity is about
20\% lower than under similar conditions in air. A similar propagation
velocity is found when photo-ionization is turned off and a background
ionization level of 10$^{7}$\,cm$^{-3}$ for positive and negative
ions is assumed. However, streamers with strong photo-ionization as
in air are thicker than with the weak photo-ionization of nitrogen
with 1 ppm oxygen.

In~\cite{Wormeester2010} we find that in air, at streamer repetition
rates up to 1\,kHz, the streamer propagation is dominated by photo-ionization,
while in high purity nitrogen, background ionization already can dominate
at a pulse repetition rate of 1\,Hz. Finally, we see that our short
streamers already branch in {}``pure\textquotedblright{} nitrogen,
but not yet in air under otherwise identical conditions.

We finally remark that all above simulations are performed by approximating
the charged particles by their densities. But in~\cite{Wormeester2011},
we have argued that the feathery structures in high purity nitrogen
are probably due to single electrons creating ionization avalanches.
Resolving the avalanche density therefore requires an approach resolving
single electrons. Such methods were recently developed by~\cite{Li2009,Li2011,Li2011a,Chanrion2010,Luque2011a}.

\subsection{Discussion and approach of the paper}

We here investigate the importance of background ionization for streamer
inception, propagation and structure by artificially increasing this
background ionization. There are a few methods to increase ionization
levels in gasses. The simplest method is to use (corona) discharges.
The effects of leftover ionization from these prior discharges on
a new discharge can be investigated by changing the repetition rate
of the voltage pulses. We have measured the effect of repetition rates
between 0.01 and 10\,Hz. The results of these measurements and related
topics are discussed in section~\ref{sec:Effects-of-repetition}.
If the repetition rate has a significant influence on streamer propagation,
one could expect that streamers in subsequent discharges follow the
pre-ionized paths of previous discharges. In section~\ref{sec:Repetition-of-streamer}
we investigate this question both experimentally and theoretically.

Another means to artificially enhance background ionization in a gas
is to add radioactive compounds. The alpha and/or beta particles created
in the radioactive decay can ionize many more gas molecules, thereby
increasing the background ionization. The advantage of this method
is that the resulting ionization level on average is homogeneous in
space and time. Disadvantages are that the gas mixtures with radioactive
admixtures are expensive and require special handling. We have performed
measurements on streamer discharges in nitrogen with a small addition
of radioactive krypton-85. The results of these measurements are discussed
in section~\ref{sec:Addition-of-radioactive-Kr85}. By comparing
them with the effects of repetition rate, a consistent picture of
the effects of background ionization levels on positive streamers
is formed.

There are more methods to increase background ionization. For example
irradiation with ultraviolet light (UV) or x-rays. However, radiation
that is still transmitted by the windows of our vessel (wavelengths
above $\sim$170\,nm) is not able to directly ionize nitrogen or
oxygen molecules. It can be used to liberate electrons from electrodes
and walls or from impurities, but this is of limited use in our large
vessel as it would lead to a very inhomogeneous distribution of ionization
density. At higher intensities it is possible to ionize molecules
or atoms with two or more photons but these high intensities make
it impossible to produce homogeneous ionization levels throughout
the vessel.

With special windows, it would be possible to use UV or x-ray radiation
of lower wavelengths. However, both the installation of such windows
as well as the generation of enough radiation of the right wavelengths
would take a lot of effort. Furthermore, this would still result in
a quite inhomogeneous distribution of background ionization with an
unknown density. Takahashi \emph{et al.}~\cite{Takahashi2011} have
used a KrF laser to increase the background ionization in pure argon
as was discussed above. They use a relatively small beam (1\,cm diameter),
and therefore they have enough intensity to achieve two-photon ionization
in part of their gas volume, while one-photon ionization is impossible
at the wavelength they use.

More exotic methods like electron or ion beams to directly inject
charges into the volume have to be discarded for similar reasons.

Besides the nitrogen-krypton mixture discussed above, we use pure
nitrogen (in two purity levels) and artificial air in our experiments.
(Artificial) air is the gas mixture that is mostly used for streamer
research and therefore makes a good benchmark, while the low oxygen
levels in pure nitrogen suppress photo-ionization and therefore give
more insight into the effects of background ionization. To better
understand our experimental results, we have calculated the expected
values of background ionization levels for various relevant conditions.
The methods and results of these calculations will be given in section~\ref{sec:Calculating-background-ionization}.

\section{Ionization estimates and experimental conditions}

\subsection{\label{sec:Calculating-background-ionization}Calculating background
ionization levels}

Here we show the mechanisms that determine background ionization and
give some rough estimates of the expected recombination times and
ionization levels under different conditions. Especially the recombination
times are important as they determine the ionization levels from previous
discharges, as well as the equilibrium level when a constant electron
source is present (like in the case of a radioactive admixture).

\subsubsection{Electron attachment}

As streamer discharges produce high (local) ionization levels, they
can increase the background ionization for subsequent discharges.
In gas mixtures with high oxygen concentrations, electrons quickly
attach to oxygen molecules~\cite{Kossyi1992,Pancheshnyi2005}:
\begin{eqnarray}
\mathrm{e+O_{2}+O_{2}} & \rightarrow & \mathrm{O_{2}^{-}+O_{2}}\label{eq:Att-1}\\
 &  & k_{att1}=2\cdot10^{-30}\:\mathrm{cm^{6}\, s^{-1}}\nonumber
\end{eqnarray}
 or, in gasses with low oxygen concentrations:
\begin{eqnarray}
\mathrm{e+O_{2}+N_{2}} & \rightarrow & \mathrm{O_{2}^{-}+N_{2}}\label{eq:Att-2}\\
 &  & k_{att2}=8\cdot10^{-32}\:\mathrm{cm^{6}\, s^{-1}.}\nonumber
\end{eqnarray}
This isbased on three body attachment, which is only valid for low
fields and high gas densities ($E/n_{0}<5$\,Td~\cite{Hagelaar2005,Dujko2011}).
Inside a streamer head fields will be significantly above this level,
but after the streamer phase has passed, fields will be below it.
\\
The free electron density $n_{e}$ will decrease as function of time
as
\begin{eqnarray}
\partial_{t}n_{e}= & - & k_{att1}\cdot n_{e}\cdot\left[\mathrm{O_{2}}\right]^{2}\label{eq:Attachment-differential}\\
 & - & k_{att2}\cdot n_{e}\cdot\left[\mathrm{O_{2}}\right]\cdot\left[\mathrm{N_{2}}\right]\nonumber
\end{eqnarray}
 if diffusion and drift are neglected. The attachment time~$t_{att}$
(the e-folding time of the exponential decay) is determined by~(\ref{eq:Attachment-differential})
as
\begin{equation}
t_{att1}=\left(k_{att1}\cdot[\mathrm{O_{2}}]^{2}\right)^{-1}\label{eq:Attachment-time-1}
\end{equation}
 for air, or for oxygen concentrations below a few percent:
\begin{equation}
t_{att2}=\left(k_{att2}\cdot[\mathrm{O_{2}}]\cdot[\mathrm{N_{2}}]\right)^{-1}.\label{eq:Attachment-time-2}
\end{equation}
The attachment time can vary significantly depending on oxygen content
and pressure: in air at standard temperature and pressure, $t_{att1}$
is about 20\,ns while in pure nitrogen with about 0.1\,ppm of oxygen
contamination at 200\,mbar, $t_{att2}$ is about 5\,s. The rates
used here are based on cold gas with no electric field present. Therefore
the ratio between free electrons and negative ions will be many orders
higher in air than in this \textquotedbl{}pure\textquotedbl{} nitrogen.

\subsubsection{Recombination reactions}

In air, after electron attachment, two- and three-body ion-ion dissociative
recombination determines the rate of plasma decay (Pancheshnyi~\cite{Pancheshnyi2005}):
\begin{eqnarray}
\mathrm{O_{2}^{-}+O_{4}^{+}} & \rightarrow & \mathrm{O_{2}+O_{2}+O_{2}}\label{eq:Recombination-1}\\
 &  & k_{rec1}=10^{-7}\,\mathrm{cm^{3}\, s^{-1}}\nonumber \\
\mathrm{O_{2}^{-}+O_{4}^{+}+M} & \rightarrow & \mathrm{O_{2}+O_{2}+O_{2}+M}\label{eq:Recombination-2}\\
 &  & k_{rec2}=2\cdot10^{-25}\,\mathrm{cm^{6}\, s^{-1}}\nonumber
\end{eqnarray}
 where M denotes either a neutral oxygen or a neutral nitrogen molecule.
The $\mathrm{O_{4}^{+}}$ in these reactions is created quickly after
the discharge from reactions of $\mathrm{N_{2}^{+}}$ to $\mathrm{N_{4}^{+}}$,
$\mathrm{O_{2}^{+}}$ and finally $\mathrm{O_{4}^{+}}$, see e.g.~\cite{Aleksandrov1999}.
The effective recombination rate for oxygen:nitrogen mixtures is therefore
\begin{equation}
k_{rec-air}\approx k_{rec1}+k_{rec2}\cdot[\mathrm{M}].\label{eq:krec}
\end{equation}

For pure nitrogen (i.e., without electron attachment), we use $k_{rec-N_{2}}=5\cdot10^{-7}\,\mathrm{cm^{3}\, s^{-1}}$,
which, at room temperature, is equal to $k_{rec-air}$ at a pressure
of about 80\,mbar ($[\mathrm{M}]=2\cdot10^{18}\,\mathrm{cm}^{-3}$).
This effective recombination rate includes many different reactions.

\subsubsection{Homogeneous recombination\label{sub:Homogeneous-recombination}}

If spatial inhomogeneity, diffusion, and the drift of charged particles
are neglected, the ionization density $n$ in a neutral plasma with
$n=n_{+}=n_{-}$ decreases as function of time $t$ as
\begin{equation}
\partial_{t}n=-k_{rec}\cdot n^{2};\label{eq:Recombination-diff-equation}
\end{equation}
here $t$ is time, and $k_{rec}$ is the effective recombination rate
for given gas composition and density, as described above. Equation~(\ref{eq:Recombination-diff-equation})
is solved by
\begin{equation}
n(t)=\frac{n(0)}{1+n(0)\; k_{rec}t},\label{eq:ChargeDenseTimeComplete}
\end{equation}
where $n(0)$ is the ionization density at time 0; for a streamer
under standard conditions it is about 10$^{13}$ to 10$^{14}$\,cm$^{-3}$.
For all practical values of $k_{rec}$, for typical values of $n(0)$
in a fresh streamer and for timescales above milliseconds this equation
is well approximated by
\begin{equation}
n(t)=\frac{1}{k_{rec}t},\label{eq:ChargeDenseTime}
\end{equation}
i.e., it is independent of the initial ionization density. This means
that 1~second after a discharge, the ionization density in air at
atmospheric pressure will be roughly $2\cdot10^{5}\,\mathrm{cm^{-3}}$,
when homogeneity is assumed. At 0.1\,s and 10\,s after a discharge,
the ionization density will be ten times as high or low respectively.
At other pressures, this number will be different, according to equation~(\ref{eq:krec}).
Therefore, in air, at 200 and 25\,mbar, the ionization density 1~second
after a discharge will be $9\cdot10^{5}$ and $4\cdot10^{6}\,\mathrm{cm^{-3}}$
respectively.

In pure nitrogen (with less than 1\,ppm oxygen concentration), the
attachment time will always be much longer than the recombination
time. Therefore, electrons will mostly recombine before they attach
to oxygen. This means that the negative charges in pure nitrogen will
mostly consist of free electrons, while in air they are mostly negative
ions.\\

Note that the reactions above are relevant for nitrogen-oxygen mixtures
at pressures close to atmospheric. At much lower and higher pressures
and in other gasses, other mechanisms start to dominate the recombination
reactions. For example, at pressures relevant for sprite discharges
(below 1\,mbar), dissociative attachment and recombination become
very important loss mechanisms~\cite{Sentman2008}. Furthermore,
we assume that the discharge occurs in an environment where bulk gas
recombination is the major ionization loss process. This means that
the vessel or room dimensions have to be sufficiently large to neglect
recombination at the wall, and that there should be no (strong) gas
flow which can refresh the gas between repetitive pulses.

\subsubsection{Recombination and diffusion}

The calculation above assumes that the ionization is distributed homogeneously,
and that it only changes by recombination. However, as the ionization
by streamers is produced in narrow channels, diffusion will have an
influence on ionization levels as well. When we include diffusion,
equation~(\ref{eq:Recombination-diff-equation}) changes into
\begin{equation}
\partial_{t}n=D_{ion}\cdot\nabla^{2}n-k_{rec}\cdot n^{2},\label{eq:Recombination-diffusion-diff-equation}
\end{equation}
 with $D_{ion}\approx5\cdot10^{-2}\,\mathrm{cm^{2}\cdot s^{-1}}$
the diffusion coefficient under standard conditions in air \cite{Pancheshnyi2005,Raizer1991,Snuggs1971}.
The diffusion coefficient scales inversely with pressure. In Section
4, we will evaluate equation~(\ref{eq:Recombination-diffusion-diff-equation})
and show that the more localized ionization distribution in the trails
of the streamers will create lower final ionization densities than
obtained from the homogeneous estimate~(\ref{eq:ChargeDenseTime}).

\subsubsection{Constant ionization sources}

There can be a constant source of ionization~$S$ in a gas; this
source can be natural radioactivity, cosmic rays or a radioactive
admixture. Assuming again homogeneity, equation~(\ref{eq:Recombination-diffusion-diff-equation})
has to be extended to
\begin{equation}
\partial_{t}n=S-k_{rec}\cdot n^{2}.\label{eq:Recombination-diff-equation-source}
\end{equation}
In steady state ($\partial_{t}n=0$), the production rate equals the
recombination rate, and the equilibrium ionization density is
\begin{equation}
n=\sqrt{\frac{S}{k_{rec}}}.\label{eq:ChargeDenseFixed}
\end{equation}
Again, the major negative charge carrier of the (quasi-neutral) ionization
is very dependent on oxygen concentrations: at high concentrations
(e.g., air) it is dominated by $\mathrm{O}_{2}^{-}$ while in \textquotedbl{}pure\textquotedbl{}
nitrogen it is dominated by free electrons.

\subsection{Experimental conditions}

All experiments presented here have been performed on positive streamers
in a $50\times50$\,cm (diameter~$\times$~height) vacuum vessel
with a 160\,mm separation point plane geometry. This vessel is specifically
designed for operation with high purity gasses at pressures between
25 and 1000\,mbar. More details about this vessel can be found in~\cite{Nijdam2010}.

\begin{table}
\caption{\label{tab:Gas-Purity}Gas impurity levels of relevant gasses as provided
by the gas supplier. Impurity levels are given as upper limits, except
for $^{85}$Kr, which is dosed on purpose in the nitrogen/krypton
mixture. Levels are in parts per million (ppm), n.s. indicates that
no number is specified by the supplier.}

\begin{tabular}{|l|c|c|c|c|c|c|c|c|}
\hline
Gas  & N$_{2}$  & O$_{2}$  & $^{85}$Kr  & Ar  & H$_{2}$O  & CO$_{2}$  & CO  & $\mathrm{C{}_{n}H_{m}}$\tabularnewline
\hline
\hline
Nitrogen 6.0  & $\approx$100\%  & 0.2  & n.s.  & n.s.  & 0.5  & 0.1  & 0.1  & 0.1\tabularnewline
\hline
Nitrogen 7.0  & $\approx$100\%  & 0.03  & n.s.  & n.s.  & 0.05  & 0.03  & 0.03  & 0.03\tabularnewline
\hline
Artificial air  & $\approx$80\%  & $\approx$20\%  & n.s.  & 0.2  & 0.5  & 0.1  & 0.1  & 0.1\tabularnewline
\hline
Nitrogen/krypton  & $\approx$100\%  & 0.5  & 0.0099  & 5  & 0.5  & 1  & 1  & 0.5\tabularnewline
\hline
\end{tabular}
\end{table}

For our experiments we have used four different gasses: pure nitrogen
in two qualities, artificial air and a mixture of (somewhat less)
pure nitrogen with a small addition of radioactive krypton-85 (see
section~\ref{sec:Addition-of-radioactive-Kr85} for more details).
Specifications of the gas mixtures according to the suppliers are
given in table~\ref{tab:Gas-Purity}. During use, the gas inside
the set-up is flushed with such a flow rate that all gas is replaced
every 25~minutes. All experiments are performed at room temperature.
We have not found any significant differences in streamer properties
(diameter, velocity, morphology) between the two different qualities
of pure nitrogen.

To create high voltage pulses we have used two power supplies, the
so-called C-supply and the Blumlein pulser. The C-supply consists
of a capacitor that is discharged by means of a sparkgap and thereby
creates a voltage pulse with exponential rise and decay. Risetimes
of the C-supply are typically of the order 100\,ns and decay times
are of order 5\,\textmu{}s in the presented measurements. More details
about the C-supply can be found in~\cite{Briels2006}. The Blumlein
pulser consists of two ten meter long coaxial cables that are charged
and then discharged by means of a multiple sparkgap. This creates
a more or less rectangular voltage pulse with a length of about 130\,ns
and a risetime of about 10\,ns. This circuit is treated in more detail
in~\cite{Nijdam2010}. We use two different circuits for two reasons:
one is to compare with previous measurements performed with these
two circuits. The second reason is that they both have their own advantages:
the short pulse from the Blumlein pulser ensures that a rather constant
background field is applied during a well-defined time interval in
which the streamer propagates, and that it is turned off rapidly thereafter,
while the long pulse of the C-supply can create very thin and long
streamers.

The discharges are imaged by a Stanford Computer Optics 4QuikE ICCD
camera with a Nikkor UV 105\,mm f/4.5 lens. In the images presented
here, the original brightness is indicated by the multiplication factor
\textit{Mf}, similar to what Ono and Oda introduced in~\cite{Ono2003}
and as described by us in~\cite{Nijdam2010}. We have normalized
the \textit{Mf} value in such a way that the brightest image presented
in~\cite{Nijdam2010} has an \textit{Mf} value of 1. This has resulted
in the following equation for the \textit{Mf}\textit{\emph{ value,}}
\begin{equation}
M\! f=500\cdot\mathrm{e}^{\frac{V_{g}}{68.9\,\mathrm{V}}}\cdot\frac{1}{D^{2}\cdot\left(C_{max}-C_{min}\right)},
\end{equation}
 with $V_{g}$ the gain voltage of the camera, $D$ the inverse aperture
(e.g. 4.5 for f/4.5) and $C_{max}$ and $C_{min}$ the maximum and
minimum count values of a pixel used in the image representation.
The factor 500 comes from the normalization procedure and the camera
gain factor of 68.9\,V is a property of our camera. This calculation
is only valid for images from this specific camera together with our
set-up (lenses and windows) and a specific false-colour palette. Therefore
our exact definition is different from the one by Ono and Oda~\cite{Ono2003}
although the general concept is the same.

\label{Discussion-new-Mf}Note that we improved our fitting procedure
of the camera calibration data since publication of \cite{Nijdam2010}
and~\cite{Dubrovin2010}. Therefore the data presented there has
a somewhat different definition of the \emph{Mf} value. In the old
definition, we used a camera gain factor of 58.4\,V and a normalization
value of 99 instead of 500. This means that images with \emph{Mf}
values around 1 will be virtually the same in \cite{Nijdam2010} and
in this work, but that high \emph{Mf} values can be up to a factor
of 2 lower in this work than in \cite{Nijdam2010}.

\section{Effects of repetition rate\label{sec:Effects-of-repetition}}

Both pulsed power supplies allow repetition rates of about 10\,Hz
and slower. Therefore we have measured the effect of the repetition
rate on streamer morphology in high purity nitrogen and in artificial
air at 10, 1, 0.1 and 0.01\,Hz.

According to the estimates above, the repetition rate of 0.01\,Hz
will create a background ionization level of about $2\cdot10^{4}\,\mathrm{cm^{-3}}$
at 200\,mbar. This is close to natural background ionization levels.
Therefore we do not expect to see many differences between a completely
{}``fresh\textquotedblright{} gas fill, and a gas fill 100~seconds
after a streamer discharge. We will see that the higher repetition
rates yield measurable effects.

\subsection{General results}

\subsubsection{Inception, first discharge and subsequent discharges}

\begin{figure}
\includegraphics[width=1\columnwidth]{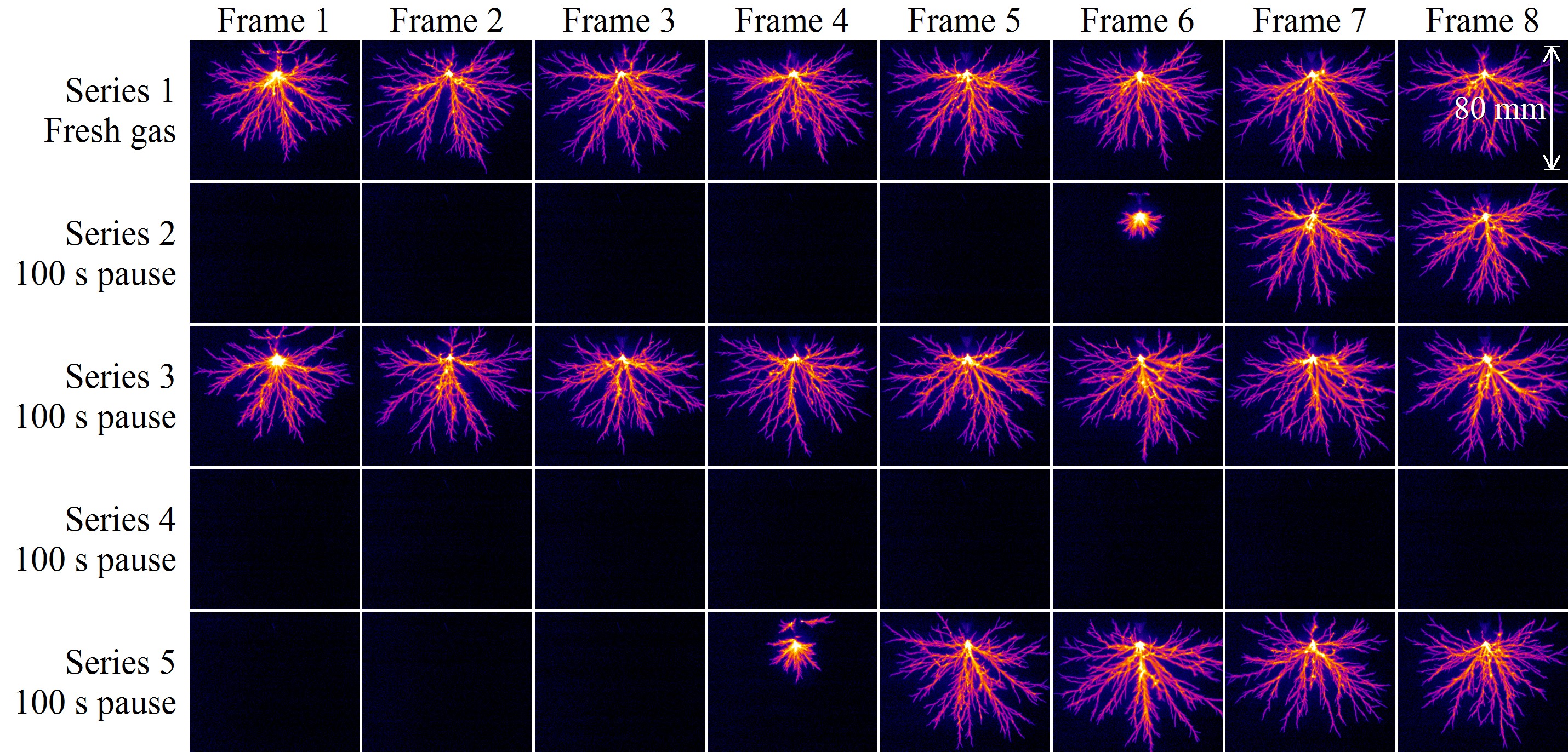}

\caption{\label{fig:FreshGasSim}Five separate series of eight consecutive
frames of streamer discharges (or attempts) in high purity nitrogen.
The series start at the first pulse of a {}``fresh\textquotedblright{}
gas filling. In series 1 this is a completely new fill of the vacuum
vessel (it was pumped down to 2\,mbar before filling). In the other
series no pulse was applied for at least 100\,s prior to frame~1.
The frames have been recorded in nitrogen 7.0 at 200\,mbar with 25\,kV
positive pulses from the Blumlein pulser at a repetition rate of 1\,Hz.
The \emph{Mf}-value of the images is about 43. The five series shown
comprise the full measurement series.}
\end{figure}

\begin{figure}

\includegraphics[width=1\columnwidth]{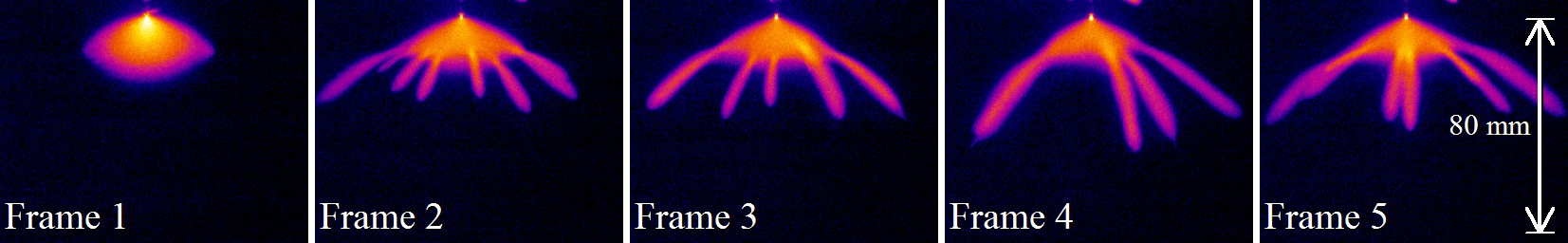}\caption{\label{fig:FreshGasSimAir}The first five frames of streamer discharges
after a 100\,s pause in artificial air. The frames have been recorded
at 200\,mbar with 25\,kV positive pulses from the Blumlein pulser
at a repetition rate of 1\,Hz. The \emph{Mf}-value of the images
is about 190.}
\end{figure}

Streamer discharges in 200\,mbar pure nitrogen with 25\,kV, 130\,ns
pulses from the Blumlein pulser often do not initiate at a pulse repetition
rate of 0.01\,Hz. Therefore we have used a different method to investigate
streamer propagation at this repetition rate: After a 100\,s pause
(no pulses), we switch the pulse source on with a repetition rate
of 1\,Hz. ICCD camera frames of the first ten to thirty pulses are
then captured and stored. In this way, the first pulse which initiates
a streamer as well as some subsequent discharges are nearly always
captured. In one case, we have taken this approach a bit further and
have used a completely fresh gas fill to start with, instead of a
100\,s pause. In this case, we first pumped the vacuum vessel down
to 2\,mbar, before refilling it to 200\,mbar. The results of these
measurements are shown in figure~\ref{fig:FreshGasSim} for pure
nitrogen and figure~\ref{fig:FreshGasSimAir} for artificial air.

In pure nitrogen we can observe that indeed the streamers do not always
initiate on the first pulse under the specified conditions. In only
two of the five attempts did the streamers initiate on the first pulse.
One of these two attempts was the case when nearly all of the gas
in the vessel was replaced (a fresh gas fill). In another attempt
(series 4), no inception at all was observed during the first ten
pulses. In general we observe an inception probability of the order
of 10\% under these conditions. In all attempts that show inception,
the morphology of the first discharge is different from subsequent
discharges. The propagation length is usually shorter and there seems
to be a denser core of the streamer tree near the tip (see also next
section).\\

In air inception is easier and (for these conditions) usually occurs
at the first pulse as is shown in figure~\ref{fig:FreshGasSimAir}.
However, we again see that the first discharge after 100\,s waiting
is different from subsequent discharges. In the first discharge, only
an inception cloud is visible, while in subsequent discharges, this
inception cloud is smaller and streamers emerge from it. It appears
that the inception cloud of the second frame is still larger than
the inception cloud of subsequent frames. We have observed the same
behaviour in all three measurement series we have recorded under these
conditions.

\subsubsection{Dependence on repetition rate}

\begin{figure}

\includegraphics[width=1\columnwidth]{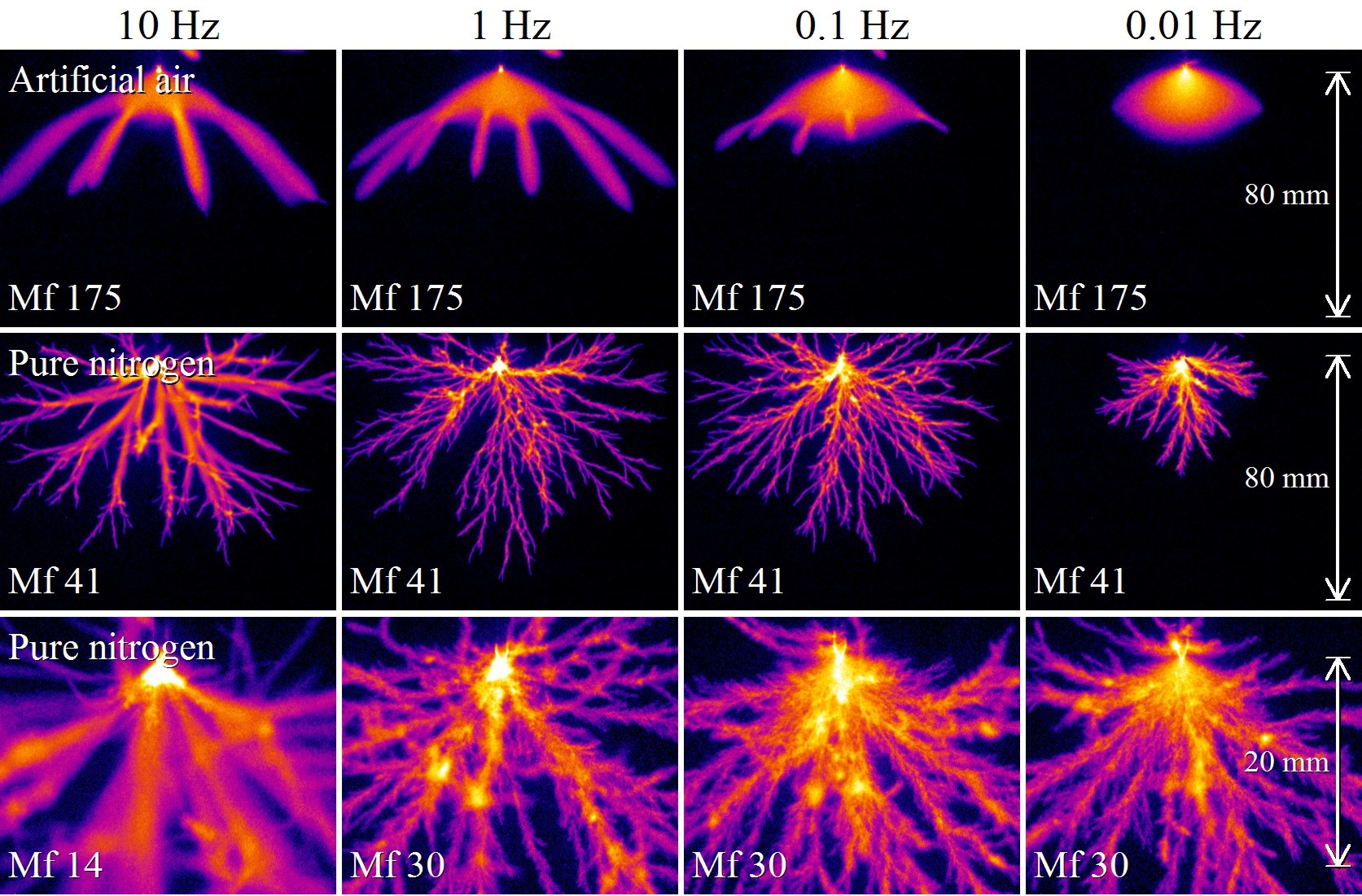}

\caption{\label{fig:Rep-freq-dep}Overview (top and middle row) and zoomed
(bottom row) images of the effects of pulse repetition rate on streamer
morphology at 200\,mbar. These images have been acquired in artificial
air and nitrogen 7.0 with the Blumlein pulser (pulse length 130\,ns)
at 25\,kV. The 0.01\,Hz images were created similarly to the ones
from figure~\ref{fig:FreshGasSim}. Shown here is the first image
with visible streamers.}
\end{figure}

We have measured the effect of the pulse repetition rate with repetition
rates from 10\,Hz down to 0.01\,Hz in pure nitrogen and in artificial
air. At least four pulses with the specified repetition rate have
occurred before capturing the images that are shown in figure~\ref{fig:Rep-freq-dep}.
Again, the measurements at 0.01\,Hz have been performed differently
as discussed above, and the first image which shows a streamer discharge
is plotted. Except for pure nitrogen at 0.01\,Hz, all conditions
had a near 100\% inception probability.

Figure~\ref{fig:Rep-freq-dep} shows that in both gasses the streamer
morphology depends on the repetition rate. In all cases, the length
of the streamers is limited by the short pulse duration ($\approx$130\,ns)
of the Blumlein pulser.

In pure nitrogen, the inception cloud is very small (as is always
the case). Here the main visible effect is that for lower repetition
rates the streamers become thinner, the number of streamers increases
and they get an increasingly feather-like structure (see also~\cite{Nijdam2010}).
The bright core that is visible in the first successful discharges
in figure~\ref{fig:FreshGasSim} and in the 0.1 and 0.01\,Hz overview
images in figure~\ref{fig:Rep-freq-dep} consists of many overlapping
feather-like streamers, as can be seen in the zoomed images in figure~\ref{fig:Rep-freq-dep}.
A more quantitative description of the feathers is given in section~\ref{sub:Effects-of-repetitionKr85}
and figure~\ref{fig:FeathersPerMM}. The total length (and therefore
the average propagation velocity) does not vary much between 10\,Hz
and 0.1\,Hz. At 0.01\,Hz, the streamers are shorter and show a large
variation in size, but this can probably be attributed to late inception.
As was shown in the previous section, streamers mostly do not initiate
on the first pulse after a 100\,s pause. Therefore we suggest that
the first streamer discharge that does initiate, does so at a random
moment during the pulse. This means that the length of the streamer
discharge at 0.01\,Hz is not a good indication for its propagation
velocity, but rather for the actual inception time within the voltage
pulse. Except for the difference in discharge length, the general
morphology is the same for 0.1 and 0.01\,Hz.

In air, we see that the inception cloud becomes larger for lower repetition
rates. Simultaneously, the streamers emitted from this cloud become
shorter and thinner. At 0.01\,Hz, there is only a hint of a streamer
coming from this cloud (on the right hand side in this example).

\subsection{Discussion\label{sub:Discussion-Effects-of-repetition}}

We have observed that at higher pulse repetition rates, there are
fewer but thicker streamer channels, both in artificial air and in
high purity nitrogen; and they branch less. This effect is more prominent
close to the electrode tip than at the outer edges of the streamer
discharge. We have not observed any difference in streamer morphology
between repetition rates of 0.1\,Hz, 0.01\,Hz and discharges in
a fresh gas fill. However, in pure nitrogen, streamers do initiate
more easily at 0.1\,Hz than at 0.01\,Hz. In artificial air, the
inception cloud is larger at lower repetition rates.

\begin{figure}

\includegraphics[width=8cm]{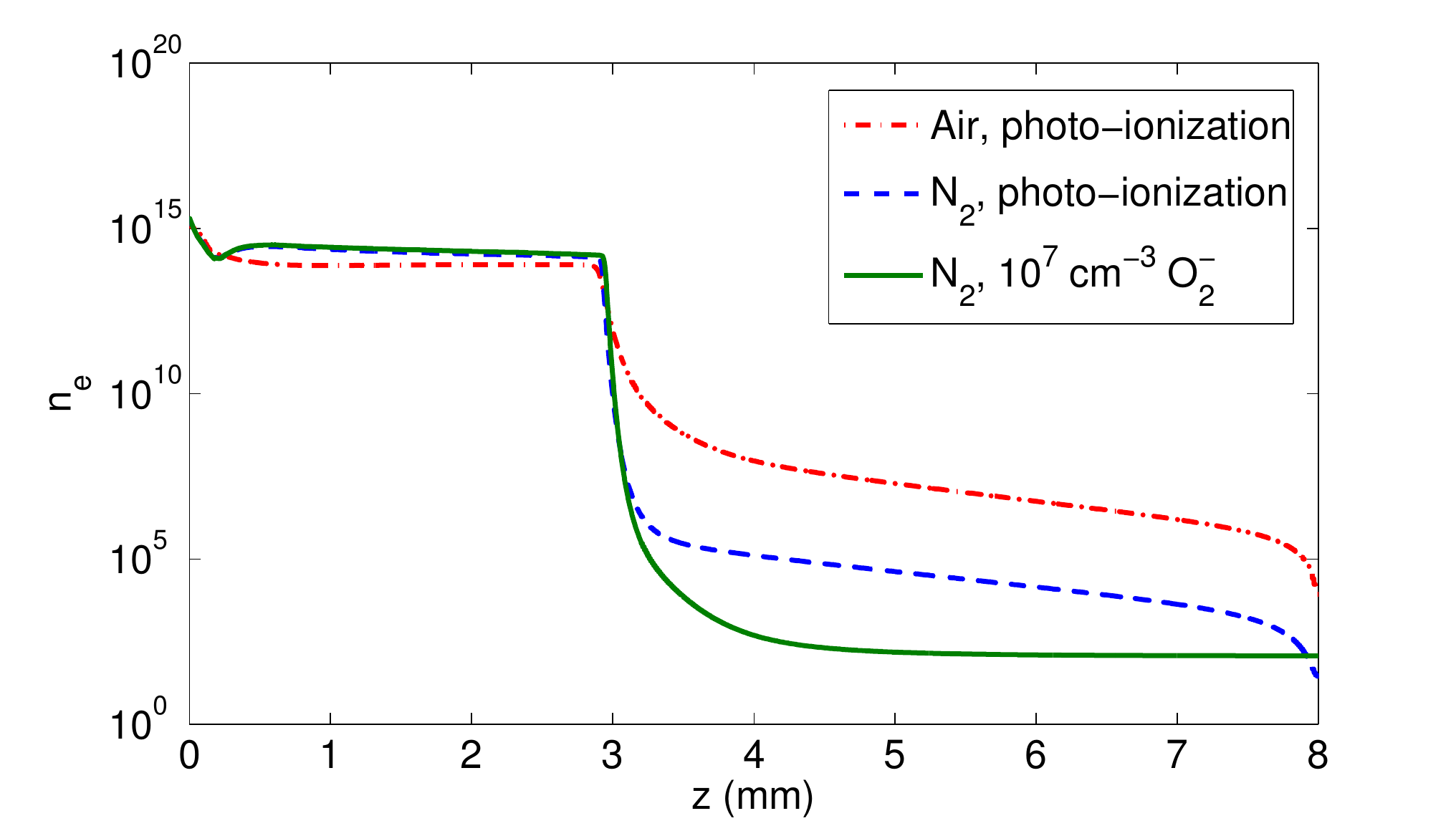}

\caption{\label{fig:Gideon-e-densities}Simulated electron densities on the
streamer axis. The streamers propagate from left to right with their
heads at about $z=3$\,mm. The solid curve shows nitrogen with a
background ionization of 10$^{7}$\,cm$^{-3}$, but without photo-ionization,
the dashed curve shows nitrogen with the photo-ionization corresponding
to an admixture of 1\,ppm oxygen, and the dashed-dotted curve shows
air with its photo-ionization corresponding. A similar figure was
discussed and published by Wormeester \emph{et al.}~\cite{Wormeester2010}.}
\end{figure}

That streamers branch less at higher repetition rates and consecutively
higher background ionization, was observed already by Pancheshnyi~\cite{Pancheshnyi2005}
and is in accordance with the observations described above and with
observations by others (e.g.~\cite{Goldman1978,Takahashi2011}).
Conceptually, this observation is consistent with the fact that streamers
are wider and more stable in air than in pure nitrogen: in air, photo-ionization
supplies a much higher level of ionization in the proximity of the
streamer head. This is illustrated in figure~\ref{fig:Gideon-e-densities}
(taken from~\cite{Wormeester2010}) that shows the electron density
on the axis of a simulated streamer in three different gases: in pure
nitrogen without photo-ionization, but with a background ionization
of 10$^{7}$\,cm$^{-3}$, in nitrogen with the photo-ionization corresponding
to an admixture of 1\,ppm oxygen, and in air with its photo-ionization.
Air has by far the highest ionization profile ahead of the streamer.
The conclusion is that a high ionization density ahead of the streamer
makes it propagate in a more stable manner, as different progression
of parts of the streamer front is averaged out by lateral diffusion
and photo-ionization if applicable.

In air, the many electrons ahead of the streamer will form so many
ionization avalanches in the active streamer ionization zone that
they cannot be distinguished visually, but appear as a thick smooth
streamer body. In contrast, in very pure nitrogen at low pulse repetition
rates, avalanches created by single electrons are so few in number,
that they can be distinguished visually according to estimates presented
in~\cite{Wormeester2011}. We there have suggested that these avalanches
can be identified with the hairs in the feathery streamer structures
close to the needle electrode as described in~\cite{Nijdam2010,Wormeester2011};
these \textquotedbl{}feathers\textquotedbl{} can also be seen in the
zoomed images at 0.1 and 0.01\,Hz in figure~\ref{fig:Rep-freq-dep}.
At higher repetition rates, the background ionization level increases,
and the streamer propagation becomes more similar to that in air.
As the streamers are denser close to the needle electrode, in a repetitive
mode the background ionization density is higher there, and the streamers
are wider and more stable close to the needle than at the outer edges
of the discharge. This can also be verified in the figures.

In very pure nitrogen in the steel vessel at repetition rates of 0.01\,Hz,
the background ionization is so low that sometimes no inception occurs
within a 25\,kV voltage pulse of 130\,ns duration. This observation
might allow conclusions on the background ionization which will be
worked out in the future.

The fact that we observe much easier inception in artificial air than
in pure nitrogen seems to contradict the results of van Veldhuizen
and Rutgers~\cite{Veldhuizen2003}. They found that ambient air,
dry air and pure nitrogen exhibit roughly similar inception probabilities
as function of voltage. However, this could be due to the purity of
their nitrogen being three orders of magnitude less than ours, as
it was specified as 4.0, while we use 7.0 nitrogen. Furthermore, they
triggered the discharge in a somewhat unsystematic manner manually
every 10 to 60 seconds~\cite{Hermans2004}.

We have argued that at low background ionization densities and low
photo-ionization, avalanches created by individual electrons can be
visible. We cannot decide at the moment whether they stay in the avalanche
phase (in that case they only run backwards towards the parent streamer
channel), or whether they eventually develop their own space charge
and become streamer branches and run forward. Only in the second case
one could speak of streamer branching. We stress that streamers require
sufficient net charge to support their propagation; this could explain
why the \textquotedbl{}feathers\textquotedbl{} stay short.

On the contrary, if streamers appear visually smooth in air or at
high repetition frequencies, one cannot conclude that the fluctuations
due to the discreteness of electrons do not play a role. Luque \textit{et
al.}~\cite{Luque2011a} recently have shown that the branching of
positive streamers in ambient air is accelerated when the discreteness
of electrons is taken into account, and that the predicted branching
rate agrees with experimental observations.

Finally, we stress an apparent paradox: If ionization density is smoother
at higher repetition frequencies, particle fluctuations should be
less and instabilities should set in later. Therefore, one would expect
that with higher repetition frequency, the inception cloud around
the needle electrode in figure~2 would become larger before it destabilizes
and breaks up into streamers, but the opposite is the case. We have
no explanation for this observation yet, except for the hypothesis
that the quite thick streamers leave some macroscopic perturbations
of the medium behind even after 1\,sec. These could be ionization
trails or thermal expansion and convection~\cite{Liu2011,Starikovskii2008}.

\section{Repetition of streamer paths\label{sec:Repetition-of-streamer}}

\subsection{General results}

\begin{figure}
\includegraphics[width=1\textwidth]{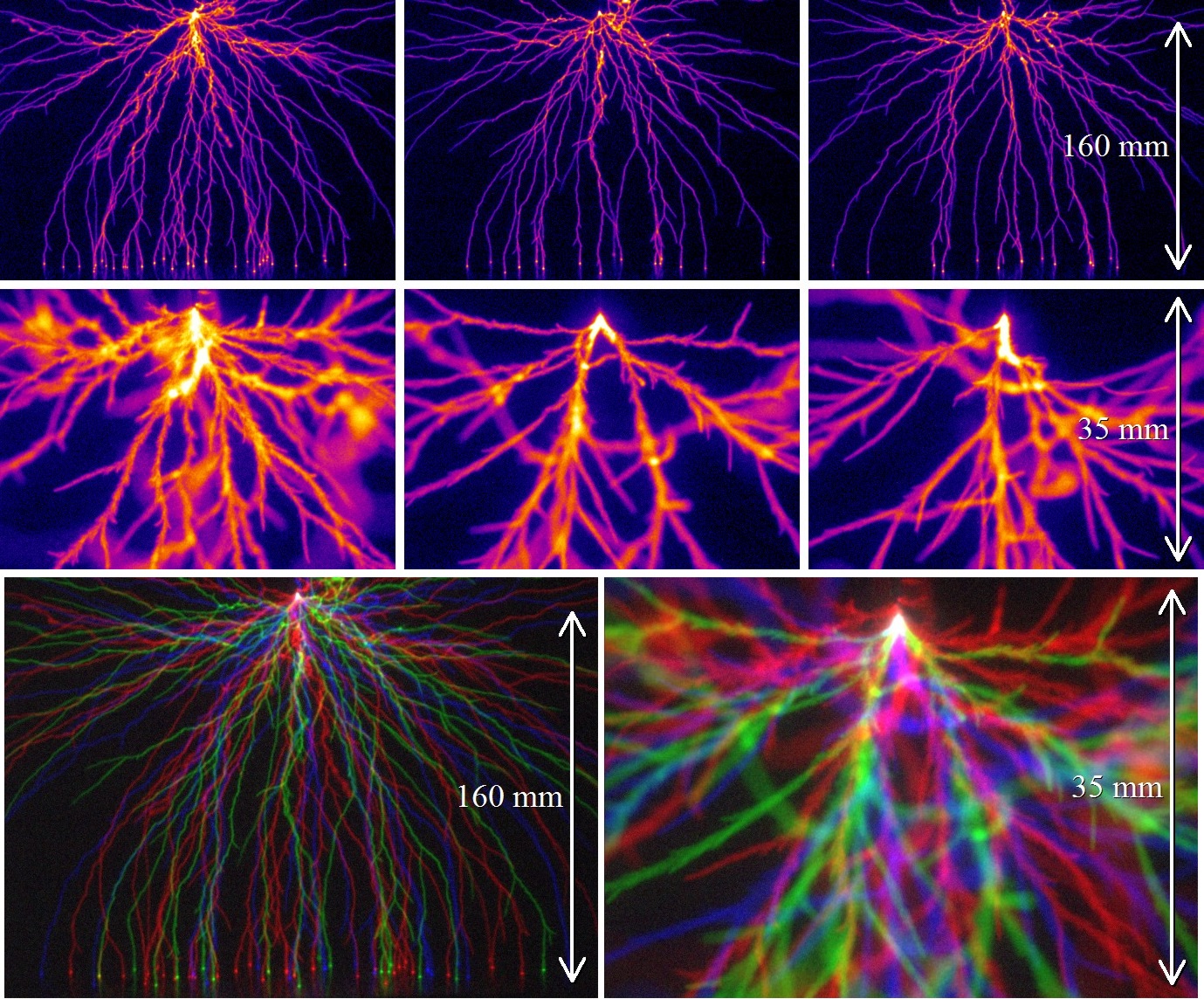}\caption{\label{fig:no-repetition-proof}Two series of three consecutive images
in nitrogen in overview (top row) and zoomed (middle row), both at
similar conditions and with 10\,Hz repetition rate. In the bottom
row, the three consecutive images from the first or second row are
coloured in red, green and blue, respectively, and overlaid. Therefore
repetitive streamer paths are rendered yellow (red + green), cyan
(green + blue) or white (red + green + blue), while single streamer
paths are rendered red, green and blue. All images have been acquired
in 200\,mbar nitrogen 6.0 with the C-supply at 15\,kV. In both series,
the first three discharge pulses after a break of about 10\,s are
shown.}
\end{figure}

If positive streamers in pure nitrogen propagate due to the background
ionization left by previous discharges, one could expect that streamers
in consecutive discharges would follow the same paths.

To check this hypothesis, images of consecutive discharges at 10\,Hz
repetition rate are plotted in figure~\ref{fig:no-repetition-proof}.
The bottom row of this figure shows overlays of three subsequent discharges
that are coloured in red, green and blue. The images show that, in
general, subsequent streamers do not follow the same path. Only some
channels (outside the crowded central area) are rendered cyan, yellow
or white, while most channels are either red, green or blue. If we
assume that the streamer paths in subsequent voltage pulses are independent
of each other, and if we take into account that the 2D images are
projections of a 3D streamer tree, then we expect similar pictures.
Therefore we can conclude that under the conditions of figure~\ref{fig:no-repetition-proof},
consecutive discharges do not follow the old streamer trail.

In the bottom half of the discharge, the average separation between
the channels in one image is over 30\,mm if we assume cylindrical
symmetry. This value is used as input for the calculations presented
below.

As said in section 2.2, the linear dimensions of the vessel are 50\,cm
and the gas is renewed every 25~minutes. The average gas flow velocity
is therefore $~$300~\textmu{}m/sec, i.e., the gas displaces on average
by 30\,\textmu{}m between voltage pulses at a repetition rate of
10\,Hz. This is 1/10 of the minimal streamer diameter in pure nitrogen
at 200 mbar according to~\cite{Nijdam2010}, and therefore negligible.

\subsection{\label{sub:Channel-repetition}Channel repetition}

We now calculate the effects of recombination and diffusion on the
ionization trail that is left by a previous streamer in pure nitrogen,
correcting the estimates given earlier in section 2.1. In this calculation
we neglect space charge and fields and assume that the plasma will
be electrically neutral soon after the external electric field has
disappeared. We remark that most ionization anyhow resides in the
electrically neutral interior of the streamer channel, and that only
a small positive charge surplus is in the streamer skin~\cite{Ratushnaya2011}.
We approximate the experimental conditions in nitrogen at 200~mbar
by assuming that the streamer leaves an ionized trail behind that
has a gaussian profile with a full width at half maximum (FWHM) of
300\,$\mu$m and a maximum ionization density of 10$^{14}$\,cm$^{-3}$.
In 200\,mbar pure nitrogen, the ion diffusion coefficient is $D_{ion}\approx0.25\,\mathrm{cm^{2}\, s^{-1}}$
and the recombination rate $k_{rec-N_{2}}\approx5\cdot10^{-7}\,\mathrm{cm^{3}\, s^{-1}}$.
As already discussed in section 2.1, most electrons stay free until
they recombine with the ions. A brief estimate of charged particle
densities and resulting electric fields shows that the electrons cannot
diffuse independently of the ions. Rather, the joined diffusion of
electrons and ions has to be modeled by ambipolar diffusion. The ambipolar
diffusion coefficient is equal to:
\begin{equation}
D_{a}=D_{ion}\left(1+\frac{T_{e}}{T_{i}}\right).\label{eq:Ambipolar-diff-coef}
\end{equation}
We assume that electron and ion temperatures are roughly equal, which
means that $D_{a}\approx0.5\,\mathrm{cm^{2}\, s^{-1}}$ and use this
number instead of $D_{i}$ in our calculations.

We now solve $n(t,r)$ in equation~(\ref{eq:Recombination-diffusion-diff-equation})
numerically. For simplicity, we assume that the streamers are distributed
in a hexagonal grid with a distance of 30\,mm to their neighbours,
which can easily be approximated by a single streamer in cylindrical
symmetry with $\frac{\partial n}{\partial r}=0$ as boundary condition
(see illustration in~\cite{Ratushnaya2011}). The initial conditions
include the gaussian streamer channel around $r=0$ together with
a homogeneous background ionization level
\begin{equation}
n(0,r)=n_{ini}+n_{channel}\cdot e^{-\frac{r^{2}}{\sigma^{2}}},\label{eq:Diff-Rec-Initial-conditions}
\end{equation}
 where $n_{ini}$ is the background ionization level, $n_{channel}$
is the maximal ionization density left in the centre of the channel,
and $\sigma$ is a measure for the width of the channel ($\sigma=\mathrm{FWHM}/2\sqrt{\ln2}$).

\begin{figure}
\subfloat{\includegraphics[width=8cm]{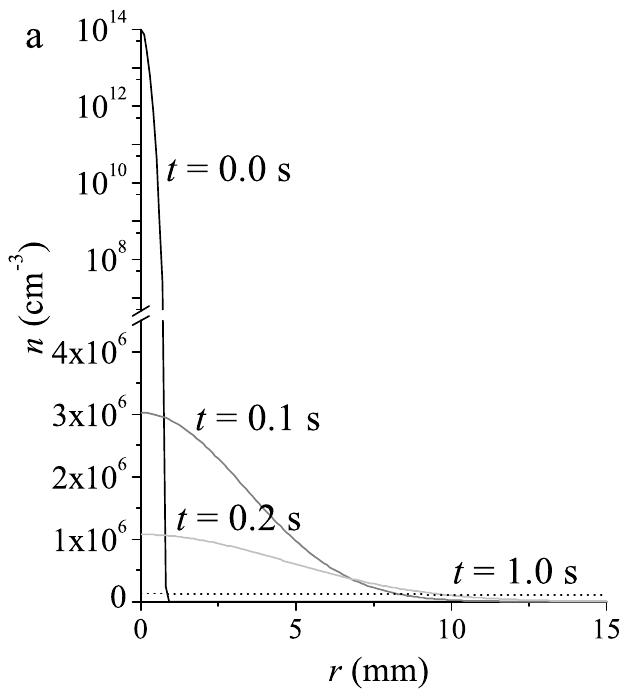}}\subfloat{\includegraphics[width=8cm]{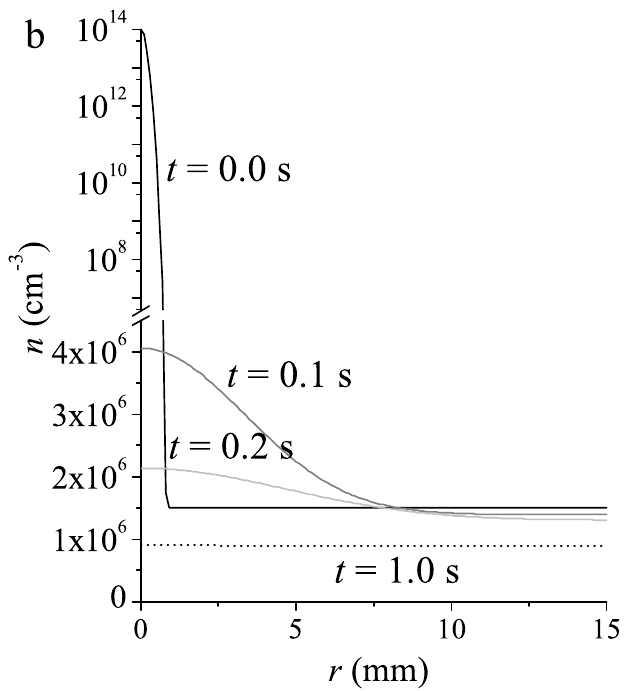}}

\caption{\label{fig:Diff-Recomb-calc}Ionization density as function of distance
$r$ to the centre of a streamer channel for different times $t$
after the discharge. Results are for pure nitrogen at 200\,mbar with
30\,mm distance to the next streamer. Initial conditions include
a background ionization level $n_{ini}=1\cdot10{}^{3}$\,cm$^{-3}$
for (a) and $n_{ini}=1.5\cdot10{}^{6}$\,cm$^{-3}$ for (b), both
with the same central gaussian streamer trail at $r=0$ ($\mathrm{FWHM}=0.3$\,mm
and $n_{channel}=10{}^{14}$\,cm$^{-3}$). Note that the vertical
axis changes from a linear to a logarithmic scale at $n=5\cdot10^{6}$\,cm$^{-3}$.}
\end{figure}

First we study the case where the initial background level is $n_{ini}=10{}^{3}$\,cm$^{-3}$,
representative for the first discharge in virgin gas (figure~\ref{fig:Diff-Recomb-calc}a).
In this case, the ionization density at $t=0.1$\,s at the centre
of the original channel ($r=0$) is about 2700 times higher than on
the edge of the domain ($r=15\,\mathrm{mm}$). This means that the
channel is still clearly present. However, its width has become quite
large ($\mathrm{FWHM}\approx$\foreignlanguage{english}{~8\,mm}),
as the figure shows. One second after the discharge the profile has
become much wider ($\mathrm{FWHM}\approx15\,\mathrm{mm}$) and the
ratio between the centre and the edge levels has decreased to about
1.2.

To determine the background ionization level after many shots, we
have searched for a value of $n_{ini}$ where the spatially averaged
calculated ionization density $n_{ave}$ at $t=0.1$\,s (averaged
over cylindrical coordinates), is equal to $n_{ini}$. We call this
the equilibrium background ionization density, $n_{eq}=n_{ave}=n_{ini}$.
Under the given conditions, $n_{eq}\approx1.5\cdot10^{6}$\,cm$^{-3}$.
Results for the ionization density distribution with $n_{eq}$ as
initial background level are given in figure~\ref{fig:Diff-Recomb-calc}b.
Here, at $t=0.1$\,s the ratio between centre and edge ionization
density is only 2.9 and the FWHM of the trail is again about 8\,mm.

From these calculations, we can conclude that in a 10\,Hz repetitive
discharge in 200~mbar nitrogen, the ionization trail left behind
by the previous discharge is quite weak. At its centre its ionization
density is less than 3 times higher than in the periphery. The ionization
density decreases by more than 7 orders of magnitude within 0.1\,s.

At lower repetition rates, the trail is even less prominent. At 1\,Hz,
the equilibrium background ionization level is about $n_{eq}\approx4.3\cdot10^{5}$\,cm$^{-3}$.
When keeping all other parameters the same as in the previous calculations,
the centre-to-edge density ratio is about 1.03. One pulse (or second)
later, the trail has virtually disappeared.

Recent 2D drift-diffusion calculations by Nikipelov \emph{et al.}~\cite{Nikipelov2011}
show that at the time-scales of interest (0.1--1\,s), the streamer
channel in air at standard temperature and pressure can even have
lower ionization densities than the periphery. They show that the
abundance of reactive (excited) species in the plasma channel leads
to higher effective recombination rates in the channel and therefore
the density (at long time-scales) can be lower than outside the channel.\\

Note that the equilibrium ionization densities $n_{eq}$ as defined
above are close to an order of magnitude lower than the ionization
levels calculated with equation~(\ref{eq:ChargeDenseTime}) for a
homogeneous initial distribution. This is because at the streamer
core the ionization density and therefore the recombination rate are
higher than in a homogeneous distribution of the same total amount
of charged species. At 200\,mbar nitrogen and 1\,Hz, the result
from equation~(\ref{eq:ChargeDenseTime}) is $n\approx2\cdot10^{6}\,\mathrm{cm^{-3}}$
while for the same conditions $n_{eq}\approx4\cdot10^{5}$\,cm$^{-3}$.
At 10\,Hz repetition rate, these values are $2\cdot10^{7}\,\mathrm{cm^{-3}}$
and $n_{eq}\approx1.5\cdot10^{6}$\,cm$^{-3}$ respectively. However,
the value of $n_{eq}$ depends on the assumptions for streamer channel
width, channel separation and maximal streamer ionization density.
For high streamer densities (due to wide channels or small channel
separation), $n_{eq}$ becomes equal to $1/\left(k_{rec}t\right)$.

For streamers in air, photo-ionization can increase ionization levels
outside of the streamer channels during streamer propagation.\\

In~\cite{Pancheshnyi2005}, Pancheshnyi also solves equation~(\ref{eq:Recombination-diffusion-diff-equation})
numerically, but not it in the radial direction of the streamer as
we do, but between planar electrodes (though he speaks about a needle-to-plane
geometry). In his approach, the ionization density inside the streamer
channel immediately after the discharge is assumed to be present homogeneously
in the whole volume. Furthermore he uses a fixed source of ionization
to account for ionization by cosmic rays and natural background ionization.

Based on our calculations, it seems doubtful whether such a homogeneous
background ionization is a reasonable approximation for the remainders
of previous discharges in air at repetition rates of 10 to 30\,kHz~\cite{Celestin2009,Bourdon2010},
since diffusion coefficients and recombination rates are similar to
those in nitrogen. However, Celestin, Bourdon \emph{et al.} use in
their models a homogeneous background ionization based on the results
of the calculations by Pancheshnyi.

\subsection{Discussion}

The experimental results presented in figure~\ref{fig:no-repetition-proof}
illustrate our general experience that within a single series of subsequent
voltage pulses with repetition rates of 10~Hz or less, streamers
follow a new path on every new voltage pulse for air, pure nitrogen
and many other gasses and gas mixtures. There is no indication that
they are in any way influenced by the shape of the previous discharge.
Only if the gap is so short that only one streamer propagates on the
symmetry axis of the experiment, subsequent streamers will follow
the same path; examples of such behavior are shown in~\cite{Briels2008b,Nijdam2010}
and many other publications. But whenever multiple channels are observed,
the streamers take a new path on each instance.

If streamers do not follow (nearly) the same path as in the previous
discharge, this can be either because there is no preference for the
previous path or because the streamers are actually repelled by the
old discharge trail. The theoretical estimates developed for our experiment
in the section above suggest that the trail of a previous discharge
is only five times more ionized than the periphery, and furthermore
smeared out over a diameter of 5\,mm due to diffusion and recombination
in 200~mbar nitrogen at a repetition rate of 10\,Hz. At lower repetition
rates, the trail will be even less visible. Apparently, the small
differences in ionization density remaining from the previous discharge
have no influence on the path of the next discharge. When an increased
recombination rate in the streamer interior is taken into account
as suggested by Nikipelov \emph{et al.}~\cite{Nikipelov2011}, the
old streamer path could even approach a lower ionization level than
the surrounding and therefore become slightly repulsive.

\section{Addition of radioactive $^{85}$Kr\label{sec:Addition-of-radioactive-Kr85}}

In order to increase the background ionization density in pure nitrogen
without changing the repetition rate, we have performed experiments
with a radioactive admixture of krypton-85.

\subsection{The admixture and its ionization rates}

Krypton-85 ($^{85}$Kr) is a radioisotope of krypton with a half-life
of 10.756~years. The most common decay (99.57\%) is by emission of
a beta particle and formation of the stable isotope $^{85}$Rb. The
emitted beta particles have a maximum energy of 687\,keV and an average
energy of 251\,keV. We will use the average energy for our calculations.
Krypton-85 has the advantage that it is an inert gas. Therefore it
is used in many applications like in leak detection~\cite{Fries1977}
or as a small admixture to argon as an ignition aid for high intensity
discharge (HID) lamps~\cite{Gallagher2010}.

\begin{figure}
\includegraphics[width=1\textwidth]{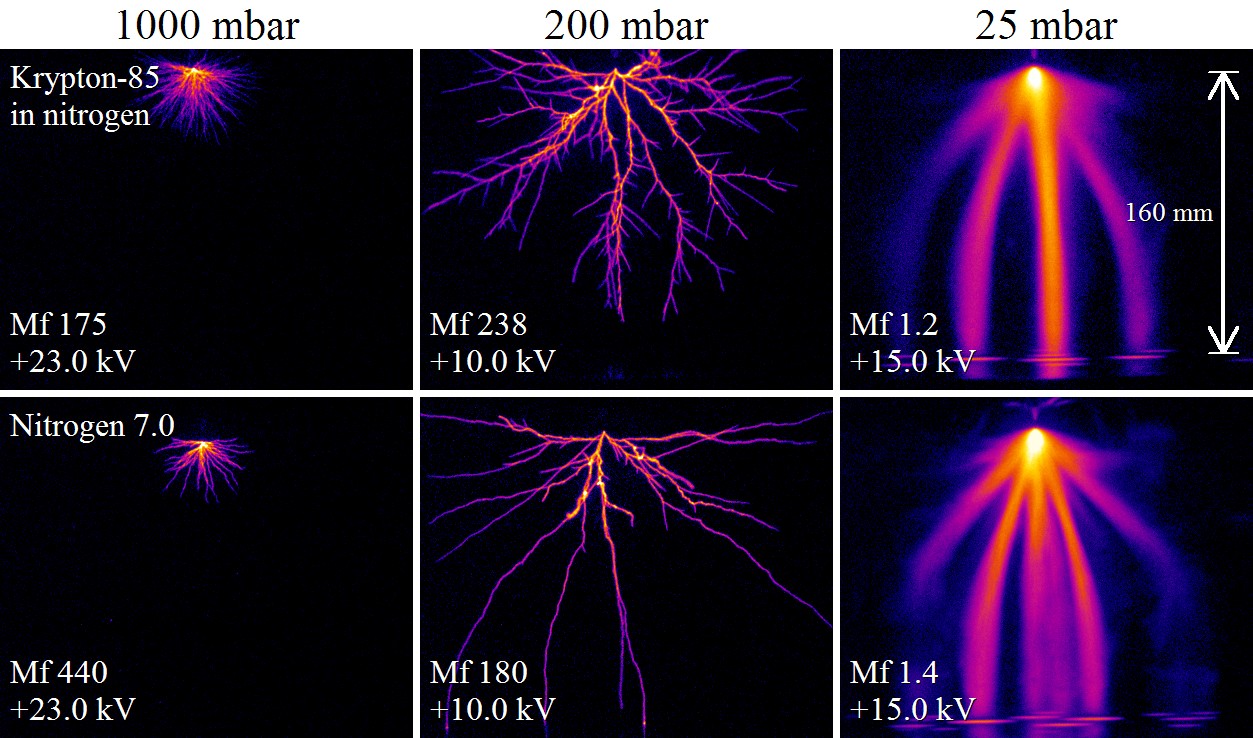}

\caption{\label{fig:Kr85-overview-C-supply}Comparison of streamer discharges
in nitrogen with an admixture of $^{85}$Kr and in nitrogen 7.0 at
1\,Hz repetition rate. The discharges are produced with the C-supply.
Discharges in other gas mixtures under the same conditions can be
found in figure 5 from~\cite{Nijdam2010}.}
\end{figure}

\label{Decay-rates}

We use a mixture of 9.9\,ppb (parts per billion) of $^{85}$Kr in
pure nitrogen. Details of the impurities in this mixture as specified
by the gas supplier are given in table~\ref{tab:Gas-Purity}. This
mixture has a specified decay rate of 500\,Bq\,cm$^{-3}$ (= 500\,kBq/litre)
at standard temperature and pressure, or about 13.5\,nCi\,cm$^{3}$
(nCi~=~nanocurie~=~37~decays~per~second). In our vessel with
a volume of about 100\,litre, this leads to about $5\cdot10^{7}$
decays per second at atmospheric pressure. The average time between
decays is 20\,ns which means that during a streamer discharge of
100\,ns to 1\,\textmu{}s duration five to fifty decays will take
place in our gas volume.\\

\begin{figure}
\includegraphics[width=1\textwidth]{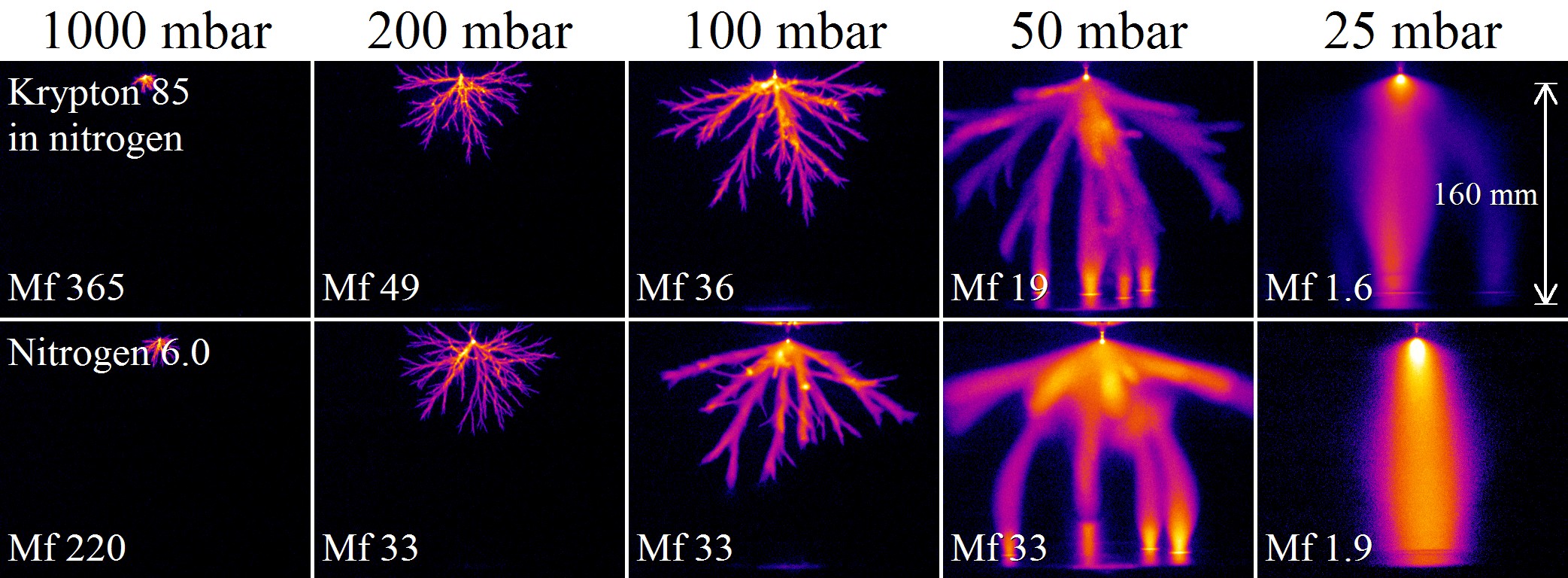}

\caption{\label{fig:Kr85-overview-Blumlein}Comparison of streamer discharges
in nitrogen with $^{85}$Kr added and in pure nitrogen. The discharges
are produced with the Blumlein pulser at about 20\,kV and 1\,Hz
repetition rate.}
\end{figure}

According to an empirical formula by Glendenin \cite{Glendenin1948,LAnnunziata2003},
the penetration length of a 251\,keV beta particle in air at atmospheric
pressure is about 50\,cm which is nearly equal to the dimensions
of our vacuum vessel. The average path length before an emitted particle
hits the wall is about 25\,cm. Therefore we assume that only half
of the energy of the particle is absorbed inside the gas.

When we assume that half of the energy loss of the particle inside
the gas is used to ionize molecular nitrogen (ionization energy 15.6\,eV),
we can estimate the electron-ion pair production at about 4000 pairs
per emitted beta particle. Combined with the decay rate given above,
this gives an ion-electron pair production rate $S$ of about $2\cdot10^{6}\:\mathrm{s^{-1}cm^{-3}}$.

We can use this production rate together with the recombination rate
for pure nitrogen $k_{rec-N_{2}}=5\cdot10^{-7}\,\mathrm{cm^{3}\, s^{-1}}$
and equation~(\ref{eq:ChargeDenseFixed}) to calculate the equilibrium
ionization density. At 1000\,mbar this gives an equilibrium ionization
density of $2\cdot10^{6}\:\mathrm{cm^{-3}}$, which is clearly higher
than the background ionization density in air, which ranges between
$10^{2}$ and $10^{4}\:\mathrm{cm^{-3}}$ depending on circumstances
(e.g. radon concentration, presence of walls, cosmic radiation shielding).

At lower pressures, the ion-electron pair production rate decreases
for two reasons: the absolute density of $^{85}$Kr is lower, and
the path length of the beta particles is longer. In other words, the
ion-pair production rate scales with $n{}_{0}^{2}$ for gas densities
$n_{0}$ when the penetration length is larger than the dimensions
of the vessel. The equilibrium ionization densities will be roughly
$4\cdot10^{5}$ and $1\cdot10^{4}\:\mathrm{cm^{-3}}$ at 200 and 25\,mbar
respectively.

\subsection{General results}

\begin{figure}
\includegraphics[width=8cm]{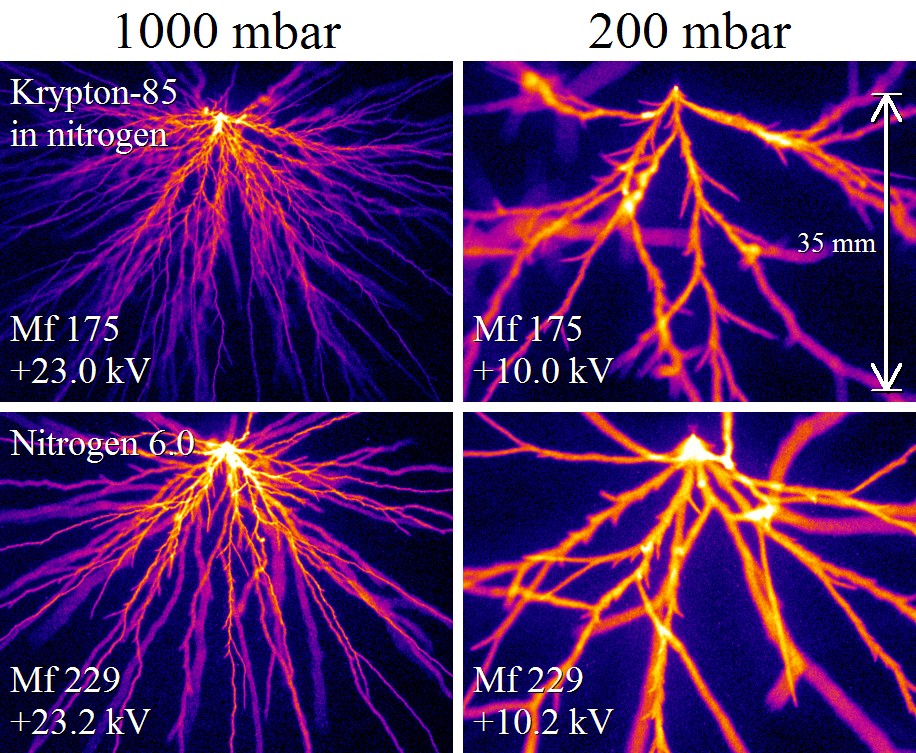}

\caption{\label{fig:Kr85-zoom-C-supply}Comparison of zoomed streamer discharge
images in nitrogen with $^{85}$Kr added and in pure nitrogen, both
at 1000 and 200 mbar around the anode tip region (see length indication
at top right image). The discharges are produced with the C-supply
at 1\,Hz repetition rate.}
\end{figure}

We have compared discharges in the krypton-85 nitrogen mixture with
discharges in pure nitrogen like discussed in~\cite{Nijdam2010}.
Results are shown in figures \ref{fig:Kr85-overview-C-supply}, \ref{fig:Kr85-overview-Blumlein}
and \ref{fig:Kr85-zoom-C-supply}. In general, the discharges are
very similar to discharges in pure nitrogen. However, there are some
clear differences: In the krypton-mixture, more branches occur. These
{}``extra\textquotedblright{} branches are longer than the protrusions
observed in pure nitrogen and pure argon, but they do not reach much
farther than about 10 streamer diameters. This is especially clear
in the 200\,mbar images from figure~\ref{fig:Kr85-overview-C-supply}
and the 50 and 100\,mbar images from figure~\ref{fig:Kr85-overview-Blumlein}.
The zoomed images at 200\,mbar from figure~\ref{fig:Kr85-zoom-C-supply}
also show longer branches in the krypton-mixture than the short protrusions
in pure nitrogen. In the same figure, at 1000\,mbar, the number of
channels is about two times higher in the krypton-85 mixture than
in pure nitrogen, which may be attributed to the same reason.

\begin{figure}
\subfloat{\includegraphics[width=8cm]{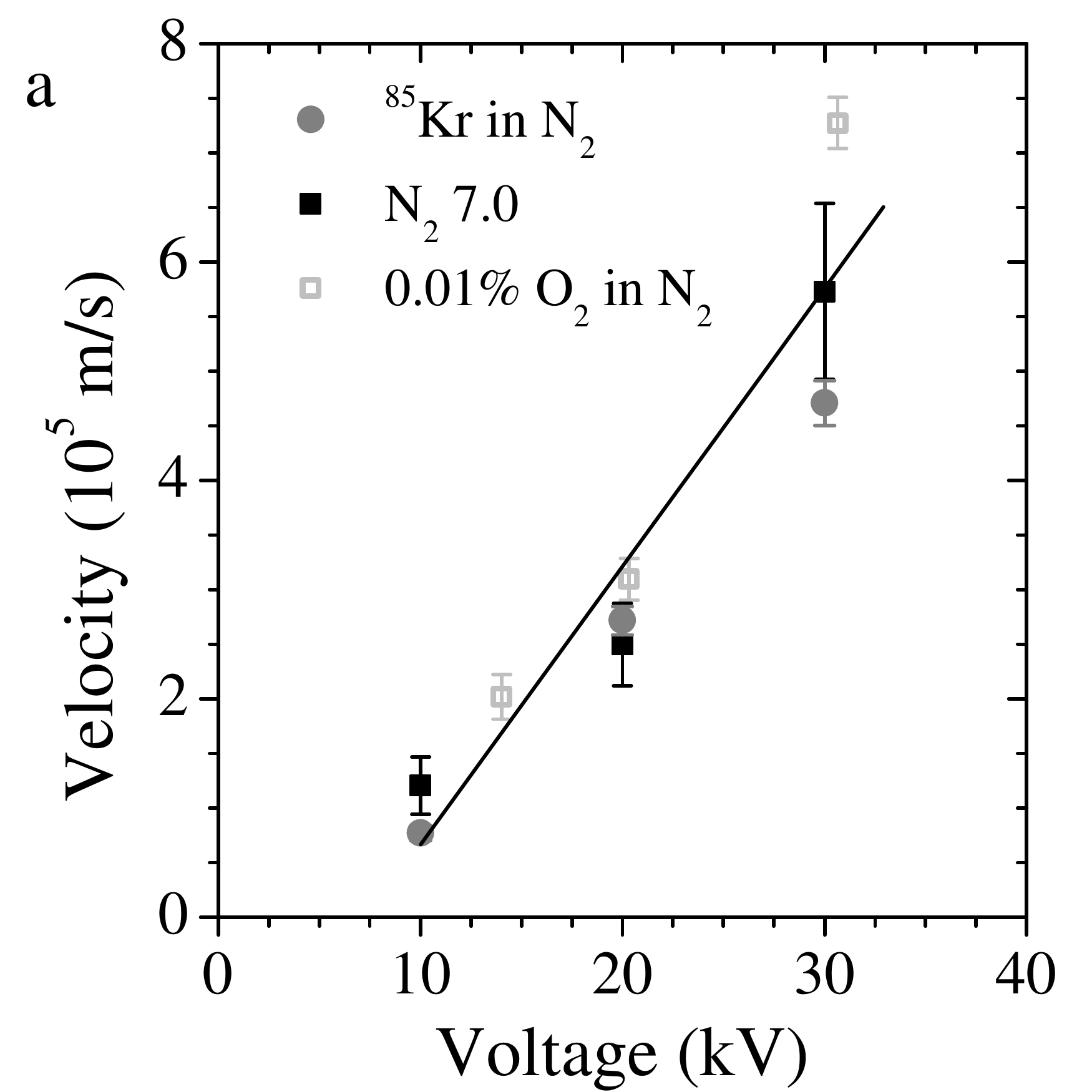}}\subfloat{\includegraphics[width=8cm]{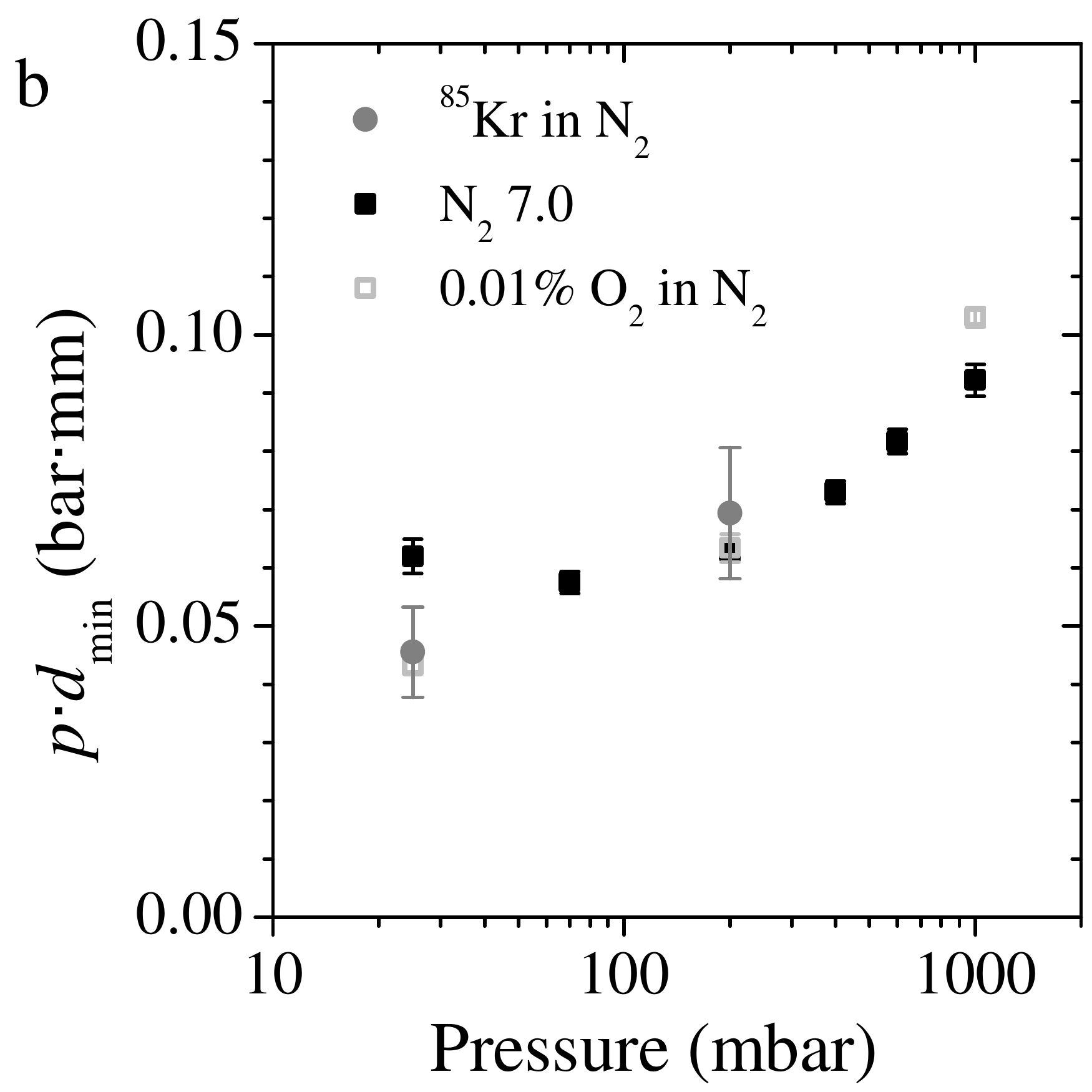}}

\caption{\label{fig:Kr85-Velocities-Diameters}Streamers created with the C-supply
at 1\,Hz repetition rate: a) velocities as a function of voltage
at 200\,mbar and b) minimal streamer diameters as a function of pressure
for the nitrogen-krypton mixture as well as for two other gasses.
The error bars indicate the sample standard deviation. The line in
a) is a linear fit through all the points from figure~13 of~\cite{Nijdam2010}.
More diameter measurements in nitrogen-oxygen mixtures are also given
in~\cite{Nijdam2010}.}
\end{figure}

However, the differences in general morphology between the krypton-mixture
and pure nitrogen have little influence on propagation velocities
and minimal streamer diameters. As is shown in figure~\ref{fig:Kr85-Velocities-Diameters},
the measured propagation velocities and minimal streamer diameters
in this mixture are very similar to values in other gas mixtures (see~\cite{Nijdam2010}).
In this case, velocities have been measured in the centre of the gap
and diameters have been determined from minimal streamers at relatively
low voltages.

\subsection{Effects of repetition rate with $^{85}$Kr\label{sub:Effects-of-repetitionKr85}}

\begin{figure}
\includegraphics[width=8cm]{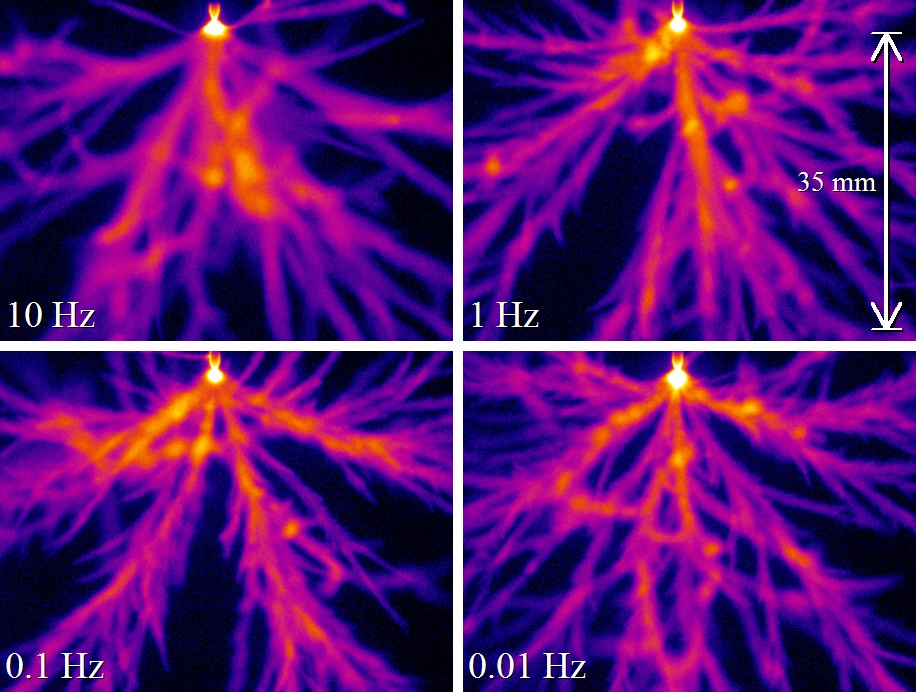}

\caption{\label{fig:Kr85-RepFreq-Zoom}Zoomed images of the effects of pulse
repetition rate on streamer morphology in the nitrogen-krypton-85
gas mixture. These images have been acquired at 200\,mbar with the
Blumlein pulser at 25~kV.}
\end{figure}

With the krypton-85 nitrogen mixture, we have performed similar experiments
as discussed in section~\ref{sec:Effects-of-repetition}. Results
of these experiments are given in figure~\ref{fig:Kr85-RepFreq-Zoom}.
Inception is very easy in this gas-mixture. In all cases a full discharge
was observed on the first pulse, also with lower voltage pulses (not
shown here). This is no surprise, as the radioactive decay of the
krypton-85 keeps supplying free electrons.

\begin{figure}
\includegraphics[width=8cm]{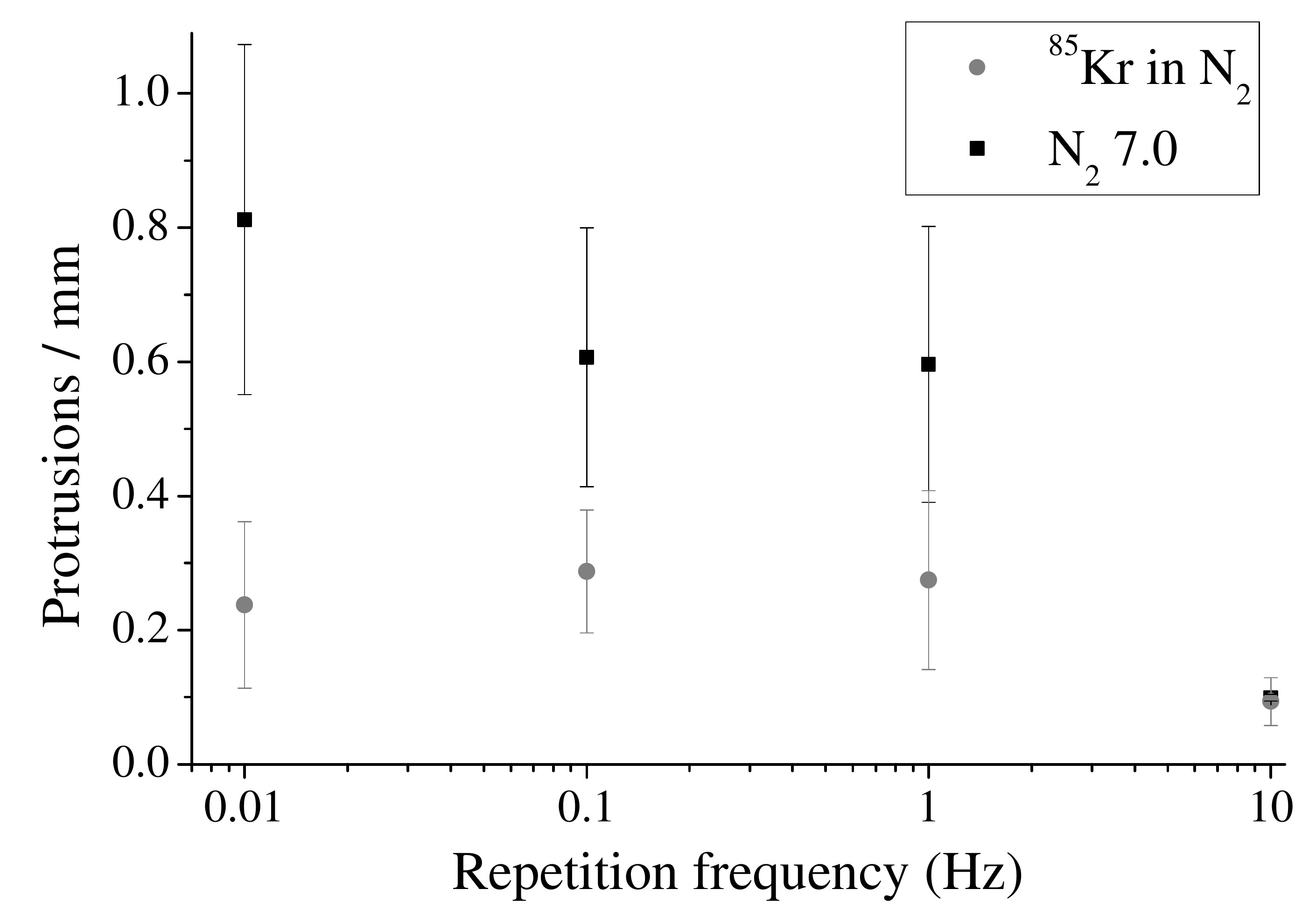}\caption{\label{fig:FeathersPerMM}Number of protrusions per millimetre as
function of repetition rate for nitrogen 7.0 and the nitrogen-krypton-mixture,
both at 200\,mbar. The same settings as in figures \ref{fig:Rep-freq-dep}
and~\ref{fig:Kr85-RepFreq-Zoom} have been used. All visible irregularities
and branches on a streamer channel are counted as protrusions (no
corrections for projection or masking by the channel are made). The
zoomed images like from figures \ref{fig:Rep-freq-dep} and~\ref{fig:Kr85-RepFreq-Zoom}
are used, with streamer channels at distances between 10 and 35\,mm
from the tip. For every point in the graph 2 to 4~images have been
used with 2 to 4 analysed streamer channels per image. The error bars
indicate the standard deviation.}
\end{figure}

In figure~\ref{fig:Kr85-RepFreq-Zoom}, the morphology of discharges
is very similar for 0.01, 0.1 and 1\,Hz pulse repetition rate. The
high density of protrusions that is observed at 0.1 and 0.01\,Hz
in pure nitrogen (see figure~\ref{fig:Rep-freq-dep}) is not visible
in the krypton-mixture. However, the thicker branches and less protrusions
at 10\,Hz that are observed in pure nitrogen are also visible in
the krypton-mixture at this repetition rate. A more quantitative approach
to this is given in figure~\ref{fig:FeathersPerMM}. Here, the number
of protrusions (hairs or branches) per millimetre of streamer channel
is plotted for the two gasses at four repetition rates. Even though
the method is quite crude (it is difficult to judge whether a structure
is a feather and there has been no correction for the projection of
the image), it gives the same general impression as the images themselves:
streamers in pure nitrogen are more feather-like than in the nitrogen-krypton-mixture,
except at 10\,Hz where both contain very few protrusions.

We can conclude from this that at 10\,Hz repetition rate, the background
ionization by previous discharges dominates over the background ionization
by radioactivity and that therefore pure nitrogen cannot be distinguished
from nitrogen with a radioactive admixture. If we use equation~(\ref{eq:ChargeDenseTime})
to calculate the leftover ionization density at 200\,mbar 1~second
after a discharge, we get $2\cdot10^{6}\,\mathrm{cm^{-3}}$. This
is about five times higher than our estimation of background ionization
caused by the krypton-85 in our gas mixture, which is $4\cdot10^{5}\:\mathrm{cm^{-3}}$
at 200\,mbar. At 10\,Hz repetition rate however, the ionization
left by previous discharges may dominate as the radioactive ionization
remains constant, while the leftover ionization increases by a factor
of 10 (compared to 1\,Hz). At repetition rates below 0.2\,Hz, the
ionization density is dominated by krypton-85. Therefore, it is no
surprise that we can no longer distinguish discharges at 1\,Hz repetition
rate from discharges at lower repetition rates, but that we still
do see differences between discharges at 1\,Hz and 10\,Hz repetition
rate.

This confirms our estimates of the effects of repetition rate and
radioactivity on background ionization levels. Together this makes
for a consistent story.

However, all ionization densities are rough estimates and can have
an uncertainty of up to one order of magnitude. Therefore, the repetition
rate at which radioactive ionization dominates over ionization from
discharge repetition has a similar uncertainty.

\subsection{Discussion}

The extra long branches observed in nitrogen with the radioactive
admixture could be due to multiple ionization events in the trace
of the fast electrons (beta-particles) from the decay of krypton-85
(see section~\ref{Decay-rates} for decay rates). However, in this
case one would expect that their orientation is fully isotropic. We
do not observe this isotropic behaviour. The extra branches are always
orientated in a small angle (less than 45\textdegree{}) to the propagation
direction of the streamer channel they are connected to. Still we
can not fully exclude this explanation as it may be possible that
only traces in the direction of the field can form avalanches.

\begin{figure}
\includegraphics[width=8cm]{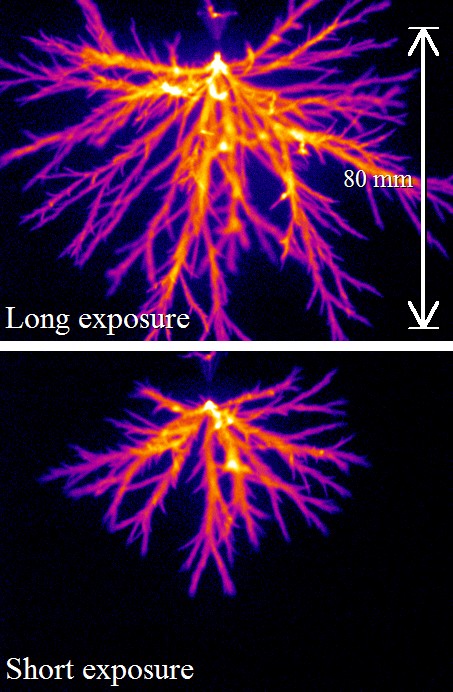}

\caption{\label{fig:Kr85-NoAvalanches}Overview images of streamer discharges
in the nitrogen-krypton-85 gas mixture at 200\,mbar made with the
Blumlein pulser at 25\,kV. The top image is made with a long exposure
so that the whole discharge evolution is visible. For the bottom image,
the camera exposure duration was reduced to roughly 65\,ns (half
of the pulse duration).}
\end{figure}

When we image a discharge in the krypton-mixture with a short exposure
time, like in figure~\ref{fig:Kr85-NoAvalanches}, we never see any
{}``disconnected\textquotedblright{} parts of a streamer channel.
If the long side channels in the krypton-mixture would be avalanches
moving towards the streamer head, one would expect to see small starting
channels visually separated from the main channels. This does not
occur in our images; we see short and long channels that are always
connected to another channel. This would indicate that the longer
side-channels in the krypton-mixture are not avalanches themselves.

Clearly, an explanation of the observed structures requires a future
investigation of the single particle dynamics of beta-particles and
ionization avalanches; it cannot be handled on the level of densities.
Steps in this direction are taken in~\cite{Li2009,Li2011,Luque2011a}.

\section{Conclusions}

We have found that the background ionization levels can have a significant
influence on streamer inception, morphology and propagation. When
background ionization levels are lower due to lower pulse repetition
rates, streamers in pure nitrogen branch more often and are more \textquotedblleft{}feather-like\textquotedblright{}
in appearance. A similar effect was seen by Takahashi \emph{et al.}~\cite{Takahashi2011}
and Mathew \emph{et al.}~\cite{Mathew2007}. when they varied background
ionization locally by UV or x-ray radiation. The higher irregularity
at lower background ionization levels is probably due to the more
stochastic nature of the discharge. Lower ionization densities will
lead to a lower avalanche density ahead of the streamer which more
easily destabilizes the streamer front and therefore creates branching
or feathers. Indeed, at the lowest ionization levels, the estimated
number of avalanches created by single background electrons approximates
the number of protrusions observed in very clean nitrogen~\cite{Nijdam2010,Wormeester2010,Wormeester2011}.

However, if all protrusions seen in the experiments were avalanches,
one would expect the number of protrusions per length to increase
with increasing background density, but figure~\ref{fig:FeathersPerMM}
shows that the opposite is true. Therefore the protrusions seen at
higher background densities are probably not pure avalanches running
towards the streamer, but they also could carry own space charge,
and possibly they have to be identified with branches that extinguish
after a short propagation length --- future theoretical research as
well as a closer inspection of the shape, length, brightness and evolution
of the protrusions needs to clarify this question. In artificial air
we observe a puzzling effect as well: the inception cloud breaks up
faster at higher repetition rates, i.e., at higher background ionization
levels, while one would expect a stabilizing effect.

Could this be because at higher repetition rates the previous discharge
has left an ionization trail behind that the next discharge will follow?
Probably not, we have not observed any tendency of streamer channels
to follow the same path as streamers in preceding voltage pulses.
This was tested, in particular, in section~\ref{sec:Repetition-of-streamer}
on streamers in 200\,mbar nitrogen with 10\,Hz repetition rate.
We found that streamers at the next pulse search new paths. This agrees
with our theoretical estimates in section~\ref{sub:Channel-repetition}
that even at the outer edges of the discharge, the ionization density
in the streamer trail is only about three times above average at pulse
repetition rates of 10\,Hz. (It should be noted though, that there
will be a substantial localized ionization trail at higher repetition
rates.)

Background ionization left by previous discharges also influences
the inception probability. We have not investigated this in depth,
but our limited measurements show that in pure nitrogen at 0.01\,Hz
repetition rate (estimated leftover ionization density $n\approx2\cdot10^{4}\,\mathrm{cm^{-3}}$
), the inception probability at voltage pulses of 25\,kV and 130\,ns
duration is in the order of 10\%, while at 0.1\,Hz ($n\approx2\cdot10^{5}\,\mathrm{cm^{-3}}$)
it is close to 100\%.

Theoretically, adding 9.9\,ppb of $^{85}$Kr to pure nitrogen at
200\,mbar in our vessel should create a background ionization level
of about $4\cdot10^{5}\,\mathrm{cm^{-3}}$. This value lies close
to the background ionization density of $10^{6}\,\mathrm{cm^{-3}}$
that was calculated in section~\ref{sub:Channel-repetition} for
a pulse repetition rate of 1\,Hz. Therefore, we expect to observe
the effect of pulse repetition on streamer morphology in the $^{85}$Kr
mixture only at repetition rates of 1\,Hz or higher. This is indeed
what the experiments show. Therefore, our rough estimates of the background
ionization levels due to pulse repetition or due to the radioactive
admixture are consistent with each other.

In the $^{85}$Kr mixture more short streamer sections appear. Whether
the shape of the ionization traces from radioactive decay can explain
this, has to be investigated in future work.

In general, our findings are in agreement with estimates and modelling
by Pancheshnyi~\cite{Pancheshnyi2005}, Bourdon \emph{et al.} \cite{Bourdon2010}
and Wormeester \emph{et al.~}\cite{Wormeester2010}. They all find
that background ionization has an influence on the structure and inception
of streamers. Our experiments with adding radioactive $^{85}$Kr to
pure nitrogen and the comparison with streamer discharges at different
repetition rates have confirmed that background ionization levels
indeed play an important role in streamer propagation, but that their
influence on propagation velocity and minimal streamer diameter is
very limited.

\ack{}{G.W. acknowledges support by STW-project 10118, part of the Netherlands'
Organization for Scientific Research NWO.}

\section*{References}

\bibliographystyle{unsrt}
\bibliography{bibliography}

\end{document}